\documentclass[final,3p,times,twocolumn]{elsarticle}

\usepackage{amssymb}
\usepackage{amsmath} 

\usepackage{subcaption}  
\usepackage{comment}     
\usepackage{acronym}     
\usepackage{booktabs}    
\usepackage{multirow}    
\usepackage{mathtools}
\usepackage{hyperref}    
\usepackage{xspace}      
\usepackage[inline]{enumitem} 
\usepackage{makecell}

\usepackage[table]{xcolor} 
\definecolor{Gray}{gray}{0.5}
\definecolor{LightGray}{gray}{0.7}

\setcellgapes{1pt}

\acrodef{acopf}[ACOPF]{alternating current optimal power flow}
\acrodef{cpu}[CPU]{central processing unit}
\acrodefplural{cpu}[CPUs]{central processing units}
\acrodef{dae}[DAE]{differential algebraic equation}
\acrodefplural{dae}[DAEs]{differential algebraic equations}
\acrodef{fgmres}[FGMRES]{flexible generalized minimal residual}
\acrodef{gpu}[GPU]{graphical processing unit}
\acrodefplural{gpu}[GPUs]{graphical processing units}
\acrodef{gmres}[GMRES]{generalized minimal residual}
\acrodef{kkt}[KKT]{Karush-Kuhn-Tucker}
\acrodef{nrbe}[NRBE]{norm-wise relative backward error}
\acrodef{nsr}[NSR]{norm of scaled residuals}
\acrodef{simd}[SIMD]{single-instruction multiple-data}
\acrodef{simt}[SIMT]{single-instruction multiple-thread}
\acrodef{gcd}[GCD]{graphics complex die}
\acrodefplural{gcd}[GCDs]{graphics complex dies}
\acrodef{olcf}[OLCF]{Oak Ridge Leadership Computing Facility}

\newcommand{\cusolver}{cuSOLVER\xspace}
\newcommand{\cusolverglu}{\texttt{cusolverGLU}\xspace}
\newcommand{\cusolverrf}{\texttt{cusolverRf}\xspace}
\newcommand{\cublas}{cuBLAS\xspace}
\newcommand{\cusparse}{cuSPARSE\xspace}
\newcommand{\exago}{ExaGO\textsuperscript{TM}\xspace}

\newcommand{\hiop}{HiOp\xspace}
\newcommand{\hykkt}{{HyKKT}\xspace}
\newcommand{\ipopt}{{Ipopt}\xspace}
\newcommand{\LDLT}{\hbox{LBL\raisebox{4pt}{\tiny\!T}}\xspace}
\newcommand{\nvidia}{NVIDIA\xspace}
\newcommand{\cuda}{CUDA\xspace}

\journal{arxiv}

\begin{document}

\begin{frontmatter}



\title{Iterative Methods in GPU-Resident Linear Solvers for Nonlinear Constrained Optimization}



\author[inst1]{Kasia \'{S}wirydowicz}
\author[inst2]{Nicholson Koukpaizan}
\author[inst2]{Maksudul Alam}
\author[inst3]{\hbox{Shaked Regev}}
\author[inst3]{Michael Saunders}
\author[inst2]{Slaven Pele\v{s}}

\affiliation[inst1]{organization={Pacific Northwest National Laboratory},
            city={Richland},
            postcode={99352}, 
            state={WA},
            country={USA}}

\affiliation[inst2]{organization={Oak Ridge National Laboratory},
            addressline={1 Bethel Valley Road}, 
            city={Oak Ridge},
            postcode={37830}, 
            state={TN},
            country={USA}}

\affiliation[inst3]{organization={Stanford University},
            addressline={450 Jane Stanford Way}, 
            city={Stanford},
            postcode={94305}, 
            state={CA},
            country={USA}}

\begin{abstract}
Linear solvers are major computational bottlenecks in a wide range of decision support and optimization computations. The challenges become even more pronounced on heterogeneous hardware, where traditional sparse numerical linear algebra methods are often inefficient. For example, methods for solving ill-conditioned linear systems have relied on conditional branching, which degrades performance on hardware accelerators such as \acp{gpu}. To improve the efficiency of solving ill-conditioned systems, our computational strategy separates computations that are efficient on \acp{gpu} from those that need to run on traditional \acp{cpu}. Our strategy maximizes the reuse of expensive \ac{cpu} computations. Iterative methods, which thus far have not been broadly used for ill-conditioned linear systems, play an important role in our approach.
In particular, we extend ideas from \cite{arioli2007note} to implement iterative refinement using inexact LU factors and \ac{fgmres}, with the aim of efficient performance on \acp{gpu}.  
We focus on solutions that are effective within broader application contexts, and discuss how early performance tests could be improved to be more predictive of the performance in a realistic environment. 
\end{abstract}








\end{frontmatter}


\section{Introduction}

There has been much recent progress on developing methods for solving ill-conditioned linear systems \cite{ina2021iterative,kardani2015parallel, gupta20133d}, such as those occurring in constrained optimization problems \cite{sanan2020ldlt,tasseff2019} or numerical integration of \acp{dae} \cite{bollhofer2020linear,baker2012enabling}. These problems are ubiquitous in decision support computations, and solving the underlying linear systems is a major part of the total computational cost.
Recent efforts  \cite{dinkelbach2021factorisation, dorto2021comparing} have focused on methods suitable for deployment on new and emerging heterogeneous hardware platforms using \acp{gpu}, which provide unprecedented computational power but introduce additional algorithmic constraints. 
While results on standalone test cases are promising, there are far fewer results describing linear solver performance within full applications. 

We focus here on linear solver performance in a realistic operating environment. We present results for \ac{gpu}-resident linear solvers we have developed and discuss the interplay between direct and iterative techniques for solving ill-conditioned linear systems. Our results show significant performance improvement with \ac{gpu}-resident linear solvers compared to a state-of-the-art baseline on \ac{cpu}, and further improvement with iterative refinement techniques developed by improving and simplifying the method proposed in 
Arioli et al.~\cite{arioli2007note}. We discuss the value of standalone tests and how much one can infer from those tests about the linear solver performance within a specific application. We also compare the performance of our refactorization linear solvers with iterative refinement against \hykkt~\cite{regev2022kkt}, a solver that implements a novel, \ac{gpu}-accelerated strategy combining direct and iterative approaches in order to improve solution efficiency.    

Our main contributions include:
\begin{itemize}
    \item A novel iterative refinement method for use with any direct linear solver. The method can adjust its solution tolerance based on ``outer loop'' nonlinear solver parameters.
    \item \ac{gpu}-resident sparse linear solvers, which combine LU refactorization and iterative refinement. The linear solvers outperform the state-of-the-art \ac{cpu} baseline when used within a power systems software stack and on standalone test linear systems.
    \item Performance analysis of \ac{gpu}-resident linear solvers and comparison between projected performance on standalone test systems and performance within the full application stack.
    \item Performance analysis of the \hykkt linear solver and comparison with LU refactorization linear solvers with iterative refinement.
\end{itemize}

The paper is organized as follows. In section \ref{sec:motivation} we detail the motivation for this research, while in section \ref{sec:description} we describe the properties of linear systems arising in our use cases. In section \ref{subsec:existing} we review state-of-the-art methods for solving these systems. We describe our iterative refinement method in section \ref{sec:ir}.
Numerical experiments and performance profiling results are presented in section \ref{sec:experiments}. In section \ref{sec:hykkt}, we analyze the performance of the \hykkt linear solver and compare it with the performance of refactorization linear solvers with iterative refinement. In section \ref{sec:conclusion} we summarize our results and discuss future research directions.

To avoid ambiguity, any reference to sparse direct solvers (or other linear solvers) is preceded by the word \emph{linear}. We use the term \emph{optimization solver} to designate optimization packages (e.g., \hiop \cite{hiop_techrep}, \ipopt \cite{wachter2006implementation}) that employ nonlinear interior methods.

\subsection{Motivation}
\label{sec:motivation}

Hardware accelerators such as \acp{gpu} are designed for high-performance computing at massive scale, but their high energy efficiency and relatively compact size also makes them suitable for applications like edge computing or embedded controls. For example, AMD's Instinct MI250X GPU has a peak performance of 47.9 TFLOPS, the equivalent of 200 high-end workstations, while drawing only 560W of power at its peak. In contrast, the cluster of high-end \ac{cpu} nodes would use 20--40kW. Hence, GPU technology allows one to have high-performance computing capability onboard a vehicle or at a distribution grid substation. Such ``portable'' computational power opens up new opportunities for the development of more sophisticated control strategies that could not be deployed on traditional microcontroller devices.

Harnessing the computational power of hardware accelerators typically requires rethinking of mathematical algorithms used in computations. For example, \acp{gpu} have one instruction sequence control per group of threads (32 for \nvidia and 64 for AMD \acp{gpu}), which means that algorithms need to be cast in terms of \ac{simt} operations in order to get maximum performance out of \acp{gpu}. Furthermore, level-1 cache is typically smaller in \acp{gpu} than in traditional \acp{cpu}, which creates challenging latency issues and makes data movement more expensive \cite{mittal_vetter2015,micikevicius2012gpu}. This, for example, makes pivoting, which is commonly used in direct linear solvers to preserve stability, prohibitively expensive on \acp{gpu}. Because of these additional hardware-imposed constraints on mathematical algorithms, many linear solvers that have performed well on traditional \ac{cpu} architectures, including distributed memory systems, cannot take advantage of hardware accelerators \cite{swirydowicz2022linear}.

Iterative linear solvers have been traditionally more effective in parallel computations~\cite{dehnavi2012parallel,desturler1991parallel} than direct linear solvers. However, the ill-conditioned nature of linear systems arising in problems like constrained optimization or numerical integration of \acp{dae} makes the use of iterative linear solvers challenging. (One essentially needs to set up preconditioners that are almost as accurate, and therefore as expensive, as direct linear solvers.)

Direct linear solvers theoretically provide exact solutions,
but in finite floating-point precision, the accuracy of the solution degrades if the system is ill-conditioned. Iterative refinement is often applied to improve the solution accuracy \cite{wilkinson1948progress,moler1967iterative}. Many theoretical developments exist on the subject (see~\cite{carson2020three} and references therein); a few authors even specifically addressed ill-conditioned linear systems~\cite{arioli2007note,carson2017}. The influence of iterative refinement on the overall performance of an application needing it is rarely studied, though. Here, we show that iterative refinement can improve the performance
of an optimization solver by reducing the interior method's
number of steps. To achieve this goal, we refine the solution of each linear system using an iterative linear solver.

\subsection{Problem description}
\label{sec:description}

As the use case for testing our linear solvers, we use a nonlinear interior method implementation deployed for \ac{acopf} analysis for transmission power grids \cite{ONeill2012}. \ac{acopf} problems have irregular and extremely sparse structure, which is typical of many complex engineered systems. As such, they are a good proxy for a broad class of decision support and optimization problems.

Interior methods have been used extensively to solve optimization problems in different engineering disciplines. Applications include model predictive controls for self-driving cars \cite{wang2020nongaussian} and aerial vehicles \cite{jerez2017forces}, power systems planning and operation \cite{ONeill2012}, and health policy strategy and development \cite{silva2013optimal,acemoglu2021lockdowns}, to mention a few. Moreover, an explosion in the use of deep learning methods for decision support has created new opportunities for deployment and further development of interior methods \cite{shlezinger2022learning}.

The appeal of interior methods is in their convergence properties where, as shown in practice, the number of steps to solution depends weakly on the size of the optimization problem \cite{gondzio2012ipm,biegler2018optimization}. This makes them suitable for solving problems with millions of variables and constraints. Most of the computational cost within interior methods comes from solving a linear system of \ac{kkt} type.  Hence, the efficiency of an interior method depends primarily on the performance of the linear solver it uses. 

Mathematically, an \ac{acopf} can be posed (without loss of generality) as a nonlinear constrained optimization problem of form
\begin{subequations}\label{eq:nlinopt}
\begin{align}
   && \min_{x\in\mathbb{R}^{n}}\ \ & f(x)    \label{eq:objective} \\
   && \text{s.t.}        \ \ & c(x) = 0,    && \label{eq:constraints} \\
   &&                        &    x \ge 0,  && \label{eq:bounds}  
\end{align}
\end{subequations}
where $f:\mathbb{R}^{n} \to \mathbb{R}$ is a possibly nonconvex objective function, and $c: \mathbb{R}^{n} \to \mathbb{R}^{m}$ defines nonlinear constraints. Interior methods enforce bound constraints (\ref{eq:bounds}) by augmenting the objective (\ref{eq:objective}) with a barrier function:
\begin{equation}
    \min_{x\in\mathbb{R}^{n}}
    f(x) - \mu\sum_{j=1}^{n} \ln{x_j},
\end{equation}
where $\mu>0$ is the barrier parameter. As $\mu$ takes a finite sequence of decreasing values,
the associated barrier subproblem is solved approximately
(with increasing accuracy) by a Newton-type method on the system of nonlinear equations 
\begin{subequations}\label{eq:nonlinearequations}
\begin{align}
  \nabla f(x) + J(x)^T \lambda - z &= 0,  \label{nonlinearequations_a} \\ 
                            c(x) &= 0,  \label{nonlinearequations_c} \\
                           X Z e &= \mu e, \label{nonlinearequations_e} 
\end{align}
\end{subequations}
where the Jacobian $J(x) \equiv \nabla c(x)$ is a sparse $m\times n$ matrix, $\lambda \in \mathbb{R}^{m}$ is a vector of Lagrange multipliers for the constraints, $z \in \mathbb{R}^{n}$ is a vector of multipliers for the bounds,  $X=\mathrm{diag}(x)$, $Z=\mathrm{diag}(z)$, $e \in \mathbb{R}^{n}$ has all elements equal to one, and $x>0$, $z>0$ is maintained throughout.

To solve (\ref{eq:nonlinearequations}), the Newton-type method solves linearized systems of the form
\begin{align} \label{eq:kktnonsymmetric}
  \begin{bmatrix}
        H   & J^T & -I
     \\ J   & 0   &  0
     \\ Z   & 0   &  X
  \end{bmatrix}
  \begin{bmatrix}
    \Delta x \\ 
    \Delta \lambda \\
    \Delta z
  \end{bmatrix} =
  \begin{bmatrix}
    \tilde{r}_{x} \\ r_{\lambda} \\ r_z
  \end{bmatrix},
\end{align}
where the right-hand side vectors are residuals of (\ref{eq:nonlinearequations}) and $H$ is the sparse $n \times n$ Hessian of the Lagrangian $L(x,\lambda)$ defined as
\begin{align}
   L(x,\lambda) &= f(x) + \lambda^T c(x),
\\ H &= \nabla^2 f(x) + \sum_{i=1}^{m} \lambda_{i} \nabla^2 c_i(x).
\end{align}
The solution to the original problem is obtained by using a continuation method to drive the barrier parameter $\mu$ close to zero and solving (\ref{eq:nonlinearequations}) at each continuation step. The \ac{kkt} matrix in (\ref{eq:kktnonsymmetric}) is singular when $\mu=0$, so the continuation method needs to stop before (\ref{eq:kktnonsymmetric}) becomes too ill-conditioned, but only after the solution of the barrier subproblem is close enough to the solution of (\ref{eq:nlinopt}). Therefore, interior methods create ill-conditioned linear systems by design. The linear solver's efficiency and its ability to deliver an accurate solution are critical for effective deployment of interior methods.

Typically, interior methods apply block Gauss elimination to (\ref{eq:kktnonsymmetric}) to solve the smaller symmetric indefinite system 
\begin{align} \label{eq:kktlinear}
\overbrace{\begin{bmatrix}
      H + D_x & J^T
     \\ J     & 0 
  \end{bmatrix}}^{K_k}
  \overbrace{\begin{bmatrix}
    \Delta x \\ 
    \Delta \lambda
  \end{bmatrix}}^{\Delta x_k}=
  \overbrace{\begin{bmatrix}
    r_{x} \\ r_{\lambda}
  \end{bmatrix}}^{r_k}
\end{align}
at each optimization solver step $k$, where $D_x=X^{-1}Z$, $r_x = \tilde{r}_x + z - \mu X^{-1} e$, and $\Delta z$ is recovered
from either $X \Delta z = r_z - Z \Delta x$
or $\Delta z = H \Delta x + J^T \Delta \lambda - \tilde{r}_x$.
For simplicity, we use $k$ to denote Newton iteration and continuation steps and suppress step indices for the matrix blocks in (\ref{eq:kktlinear}). For a more rigorous description of interior method implementation, see \cite{wachter2006implementation}.

\paragraph{Summary} Linear systems (\ref{eq:kktlinear}) are symmetric indefinite and increasingly ill-conditioned because of the nature of interior methods. In our use cases, they also have an irregular and extremely sparse structure, which is a reflection of the complex engineering systems they represent. This particular structure of the linear systems leaves little for parallel algorithms to exploit. Also, very high sparsity makes it difficult to form large dense blocks in $K_k$ by reordering rows and columns in (\ref{eq:kktlinear}). On the other hand, all linear systems $K_k \Delta x_k = r_k, ~k=1,\ldots,M$ in the sequence (\ref{eq:kktlinear}) have the same sparsity pattern. This feature can be exploited to implement an efficient linear solver.

\subsection{Existing approaches and literature review}
\label{subsec:existing}

As explained above, the systems described by \eqref{eq:kktlinear} are ill-conditioned and thus are usually solved using a sparse direct linear solver. Since the systems are classified as symmetric indefinite, a linear solver based on either LU decomposition or \LDLT decomposition can be used. In both cases, the system is solved in four steps:
\begin{itemize}
\item[1.] Symbolic decomposition: The matrix is reordered and the pattern of $L$ (or $L$ and $U$) factors is set.
\item[2.] Numeric decomposition: The factors are computed based on the pattern from the previous step.
\item[3.] Triangular solve: The factors from step 2 are used to solve the system.
\item[4.] (optional) Iterative refinement: The accuracy of the solution from step 3 is improved using the same factors.
\end{itemize}

There has been a substantial effort in the scientific community to parallelize sparse direct linear solvers \cite{schenk2000scalable, grigori2007parallel, booth2017basker}, including GPU-accelerated parallelization~\cite{peng2020glu3, Duff2020}. For systems described by (\ref{eq:kktlinear}), we run into two major stumbling blocks: (a) the systems themselves allow for very limited parallelism as they lack well-defined, reasonably sized, dense blocks of nonzeros (even with proper reordering); (b) the use of GPU accelerators exacerbates the issue, as moving around small blocks of data and deploying multiple tiny GPU kernels is extremely inefficient \cite{swirydowicz2022linear}. In \cite{swirydowicz2022linear}, the authors observed that turning on GPU acceleration within a sparse direct linear solver often results in a \textit{slowdown} compared to a \ac{cpu}-only version. Moreover, none of the tested GPU-accelerated linear solvers was faster than the \ac{cpu}-only, one-thread version of MA57 \cite{duff2004ma57}.

A possible remedy is to use \emph{refactorization}~\cite{dinkelbach2021factorisation, dorto2021comparing, swirydowicz2024gpuresident,swirydowicz2023acopf}. This term is used throughout the scientific literature in different contexts. Here 
we mean an approach in which we perform a full symbolic analysis and numerical factorization with pivoting for the first system (\ref{eq:kktlinear}), and then for each subsequent system, we only perform numerical factorization, reusing the symbolic analysis and pivot sequence obtained for the first system.
We note that for both LU-based and \LDLT-based linear solvers, symbolic factorization or reordering often dominates the computational cost. In \cite{swirydowicz2023acopf}, we used refactorization as a means to improve the GPU performance of our linear solver: the symbolic factorization was computed on a \ac{cpu}  for the first system (using the KLU solver from the SuiteSparse library \cite{davis2010algorithm}), and only a numeric factorization was done on the \ac{gpu} for subsequent systems, reusing the pivot sequence obtained for the first matrix. The resulting linear solver was the first GPU-accelerated linear solver used for \ac{acopf} analysis to outperform MA57. However, once the cost of symbolic factorization was substantially reduced, another issue showed up: triangular solves became relatively more expensive compared to factorizations.

A typical iterative refinement (Step 4) relies on Richardson-style iteration~\cite{wilkinson1948progress, moler1967iterative}. 
Let $\Delta x^{(0)}_k$ be a solution of (\ref{eq:kktlinear}) at optimization step $k$ obtained by solving $L_k U_k \Delta x^{(0)}_k = r_k$, where $K_k = L_k U_k$ (permutation and scaling of matrices are omitted for simplicity). First, the \emph{residual} $\rho^{(0)} =  r_k- K_k\Delta x^{(0)}_k$ is evaluated, preferably using twice the precision used for the matrix decomposition. If the residual norm is not small enough, $L_k$ and $U_k$ are used again to solve
\begin{equation}
    K_kd^{(0)}_k = \rho^{(0)}_k
    \label{eq:error_system}
\end{equation}
and the new solution is formed as $\Delta x^{(1)}_k= \Delta x^{(0)}_k+d^{(0)}_k$. The new residual is computed as $\rho^{(1)}_k = r_k- K_k \Delta x^{(1)}_k$. If its norm is still not small enough, the process repeats. The same idea applies if $\Delta x^{(0)}_k$ is computed using \LDLT decomposition.
Depending on the particular solver package, the linear system and the stopping criterion used, refinement can easily take $10$ or more steps, with each step adding a triangular solve with $L_k$ and $U_k$. 
One of the factorization methods used in \cite{swirydowicz2023acopf}, \cusolverglu, performs triangular solve and iterative refinement (Steps 3 and 4, respectively) in a single function without providing options to configure the refinement parameters, such as solution tolerance. Since each refinement iteration executes an additional triangular solve, the profiling results  suggest that either reducing the cost of the triangular solve or using fewer triangular solves (or both) will further improve performance.  

The Richardson-style algorithm tends to work very well for systems with moderate condition number. However, it might fail to improve the solution or even diverge for systems that are ill-conditioned \cite{arioli2007note}, especially when higher precision is not used to compute each $\rho_k$. We discuss how we address this issue in section~\ref{sec:ir}. 

An important aspect  of iterative refinement is implementing proper stopping criteria, i.e., deciding when the solution is \emph{good enough} and no further refinement is needed. An obvious stopping criterion is when the error in the solution becomes of order of the floating-point precision or smaller than a specified tolerance.
Normally, the exact solution is not known, so computing the error norm $\| \Delta x_k^{\text{true}} - \Delta x_k^{\text{computed}}\|$ is usually not possible. Thus, we need a more practical way to judge if the solution is sufficiently accurate. We observe that different implementations of sparse direct solvers may use different residual norms to ``measure'' the error, and there is no broadly accepted standard~\cite{swirydowicz2022linear}. For instance, SuperLU \cite{li05} uses the backward (component-wise) residual norm~\cite{li2003superlu_dist}, and MA57 uses the \ac{nsr}~\cite{duff2004ma57} defined as
\begin{equation}
    \text{NSR} = \frac{\|r_k-K_k \Delta x_k\|_\infty}{\|K_k\|_\infty\|\Delta x_k\|_\infty}.
\end{equation}
In theory-focused papers~\cite{carson2020three}, an estimate of the matrix norm is often required. 

This lack of standard translates to iterative refinement as well, as there does not seem to be a broadly accepted stopping condition for the iterative refinement. MA57, which implements Richardson iteration described in this section, uses a ratio-based criterion: if the ratio between old and new \ac{nsr} is small enough, the iteration stops. SuperLU repeatedly evaluates the backward residual norm to decide when to stop (using a prescribed tolerance).

\section{Iterative refinement for ill-conditioned systems}
\label{sec:ir}

\subsection{Approach}

As explained in section~\ref{subsec:existing}, the version of refactorization applied in~\cite{swirydowicz2023acopf} relies on reusing the existing symbolic factorization and performing numerical factorization with static pivoting. We generally expect the solution quality to decrease as successive systems are solved. Since our systems are ill-conditioned by design, instead of using Richardson iteration, we follow an approach similar to that of~\cite{arioli2007note} as a better choice for poorly conditioned linear systems. We solve $L_k U_k \Delta x^{(0)}_k = r_k$ and use the computed  $\Delta x^{(0)}_k$ as the initial guess for an iterative solver (\ac{fgmres}~\cite{saad1993flexible} in our case). We use $L_kU_k$ as a right preconditioner. To check for convergence, we use the relative residual condition ${\|\rho_k}^{(i)}\| <\delta_\text{tol}\|\rho_k^{(0)}\|$, where $\|\rho_k^{(i)}\|$ is the estimated (not the computed) residual norm \cite{saad1986gmres} at iteration $i$ of \ac{fgmres}. This differs from the approach in~\cite{arioli2007note}, where the convergence is checked using \ac{nrbe} and the solution is computed at every iteration. In our case, the residual norm is estimated at each iteration, and the actual solution is computed only when \ac{fgmres} is restarted and when the estimated residual norm meets the convergence criteria, which saves computational effort. We also apply the reorthogonalized classical Gram-Schmidt (CGS2) algorithm instead of modified Gram-Schmidt (as done in \cite{arioli2007note}) for added robustness and stability~\cite{Bjorck1994numerical} and also for better performance. 

It is well known that iterative refinement reduces the residual norms $\|\rho_k^{(i)}\|$
more effectively if higher precision is used to compute the residual $\rho_k^{(i)}$ for iteration $i > 1$ \cite{carson2018precision}. Here we did not use higher precision, as quad precision is not available for the \ac{gpu} devices we used, but Figure \ref{eq:error_system} shows that significant improvement was achieved even for {\it uniform precision} iterative refinement for $i = 1, 2, \ldots, 10$. In most cases one or two refinement iterations were sufficient.

We implemented the iterative refinement in CUDA/C using uniform double precision, and integrated it within \hiop. Because a long sequence of linear systems needs to be solved in our case (typically several hundred), we optimized our implementation to reuse all allocated data structures (i.e., the Krylov subspace and the buffers required by some \cuda libraries). Our implementation also features low-synchronization variants of \ac{fgmres} \cite{swirydowicz2021low, bielich2022low, yamazaki2019low}. Here, we only show results for \ac{fgmres} with CGS2 Gram-Schmidt.

Note that \ac{fgmres} can be applied in 
different ways; we found that the most practical strategy in our case is the one described in the previous paragraph. An alternative is to solve system (\ref{eq:error_system}) with an iterative linear solver, meaning the iterative solver is deployed multiple times~\cite{carson2017,arioli2009using}; see Step 5 in Algorithm 1.1 in~\cite{carson2017}. Our approach reduces the costly overhead of running the iterative solver multiple times and, in practice, performs fewer triangular solves without affecting the solution, which agrees with~\cite{arioli2007note}.

\subsection{Stopping criteria}

Iterative linear solvers like \ac{fgmres} typically use either a relative residual norm
$$
    \mathrm{RR} = \frac{\|\rho^{(i)}_k\|_2}{\|\rho^{(0)}_k\|_2} = \frac{\|r_k-K_k\Delta x^{(i)}_k\|_2}{\|r_k-K_k\Delta x^{(0)}_k\|_2}
$$
or the norm-wise relative backward error
\begin{equation}
    \label{eq:nbre}    
    \mathrm{NRBE} = \frac{\|\rho^{(i)}_k\|_2}{\|K_k\|_\infty\|\Delta x^{(i)}_k\|_2 + \|r_k\|_2}
\end{equation}
 (where the Euclidean norm of $K_k$ is sometimes used).
The \ac{nrbe} is more sensitive (especially for ill-conditioned systems) but is not practical here. In (F)GMRES, the residual norm is estimated at every iteration without explicitly computing the residual, and hence we do not have access to $\Delta x^{(i)}_k$. Evaluating $\|K_k\|_\infty$ poses another computational overhead.

Arioli and Duff in~\cite{arioli2007note} use 
$$
\mathrm{NRBE} = \frac{\|\rho^{(i)}_k\|_2}{\|K_k\|_2\|\Delta x^{(i)}_k\|_2+\|r_k\|_2}< \varepsilon
$$
as their stopping condition, where $\varepsilon$ is machine precision. In contrast, we use the relative residual RR with prescribed tolerance. While this measure is not as robust, it is far cheaper to evaluate, and allows for \emph{good enough} solutions in our test cases, as shown in section~\ref{sec:results-application}, where the number of resulting optimization solver iteration is similar to that computed with MA57. 

\subsection{Deciding whether iterative refinement is needed}

A subtle balance exists between computing a solution to the linear system that is just \emph{good enough} and performing more computations than needed for fast convergence of the optimization solver. In our approach, the decision to perform refinement or not is based on the condition $\frac{\|r_k-K_k\Delta x^{(0)}_k\|}{\|r_k\|} < \delta_\text{tol}$, where $\Delta x^{(0)}_k$ is the initial solution produced by the direct solver and $\delta_\text{tol}$ is a user-provided tolerance.  We found this strategy to be practical and quite effective for our class of problems, as it resulted in the fewest optimization solver steps compared to other tested options (see section \ref{sec:results-application}).

\section{Experiments}
\label{sec:experiments}

Evaluating the performance of a linear solver on standalone test matrices such as those from the University of Florida matrix collection \cite{davis2011university} provides helpful preliminary data to numerical linear algebra experts, but it may not be predictive of the linear solver performance within a realistic application. Better numerical experiments are needed to guide the follow-on research and prepare the linear solver for integration with the target application, especially if the application is to be deployed on heterogeneous hardware. For example, refactorization linear solvers are designed to operate within a nonlinear analysis where a sequence of linear systems (\ref{eq:kktlinear}) with the same sparsity pattern is solved. They perform expensive setup on the host device and then reuse objects created at the setup in subsequent computations, which are performed on an accelerator device (a \ac{gpu}). The expectation is that the setup cost will be amortized over subsequent linear solves. Therefore, a test on a single matrix will provide limited information on the linear solver performance in a realistic application. Another example is iterative refinement, which adds some computational overhead. Its impact on performance is seen only within a broader application scope where it can speed up the nonlinear analysis by delivering a more accurate solution at each nonlinear solver step.

We present our two-stage approach for performance testing of linear solvers. The first stage is done on a series of standalone matrices generated during a nonlinear analysis. Such a series resembles systems that the linear solver will receive in a realistic application environment. These tests allow us to compute different error norms and matrix condition numbers, which are too expensive for use in a production environment. Only when this stage is successfully completed do we move to the next (more resource-demanding) stage, where we interface the linear solver with the application software stack.

\subsection{Tests on Standalone Matrices}

The first set of tests uses a collection of \emph{standalone} matrices that were created using the \ipopt optimization engine \cite{wachter2006implementation} equipped with the MA57 linear solver. At each step of the optimization solver, a linear system was captured and saved to a file. The resulting matrices and right-hand sides are available in \cite{maack2020matrices}. For our tests, we use matrices obtained by solving an \ac{acopf} problem for synthetic grid models from the ACTIVSg series \cite{birchfield2017tamu-cases}. This allowed us to save time and effort, and interface only the most promising solver prototypes with the actual optimization engine. The test cases and their respective linear system properties are summarized in Table \ref{tab:standalone}.

\begin{table}[t] 
\caption{Characteristics of three sequences of linear systems $K_k \Delta x_k = r_k,  ~k=1,\ldots, M$  (\ref{eq:kktlinear}) used for testing linear solvers.  $N$ is the dimension of the linear system and $\mathrm{nnz}$ is the number of nonzeros in the lower triangular part (including the diagonal) of $K_k$.}
\label{tab:standalone}

\centering
\small

\begin{tabular}{lcrr} \toprule
   Name        & $M$ & $N(K_k)$ & $\textrm{nnz}(K_k)$
\\ \midrule
   ACTIVSg10k  &    20      &   238,072 &   723,720
\\ ACTIVSg25k  &    41      &   697,161 & 2,020,647
\\ ACTIVSg70k  &    17      & 1,640,411 & 4,991,820
\\ \bottomrule
\end{tabular}
\end{table}

\subsubsection{Description of the Testbed}

The testbed code was implemented in C++/\cuda (without using managed memory) and used \nvidia's \cublas and \cusparse libraries for the matrix and vector operations, and the \cusolverrf module of the cuSOLVER library for the refactorization. The first one or two matrices are factorized using KLU on the \ac{cpu}, and refactorization is used afterwards. \ac{fgmres} restarted every $10$ iterations (\ac{fgmres}(10)) is used for the iterative refinement. The results in section \ref{sec:results-standalone} only include the cost of using the direct linear solver and \ac{fgmres}(10), and omit the cost of data allocation and transfer (these operations are generally done only once per matrix series). We compare the results for KLU+\texttt{cusolverRf}+\ac{fgmres}(10) with MA57, KLU, and KLU+\cusolverglu. The last strategy has Richardson-style iterative refinement built into its solve function, which cannot be disabled. The \cusolverglu module is available through the cuSOLVER library, but is not documented or officially supported by \nvidia.

We tested three matrix series, as listed in Table \ref{tab:standalone}. With our numerical experiments, we want to answer two questions: ($i$) How much does iterative refinement improve the solution, and  ($ii$) how much does iterative refinement add to the linear solver run time?

\subsubsection{Experimental Results} \label{sec:results-standalone}

\begin{figure}[t] 
    \centering
    \includegraphics[width=\columnwidth]{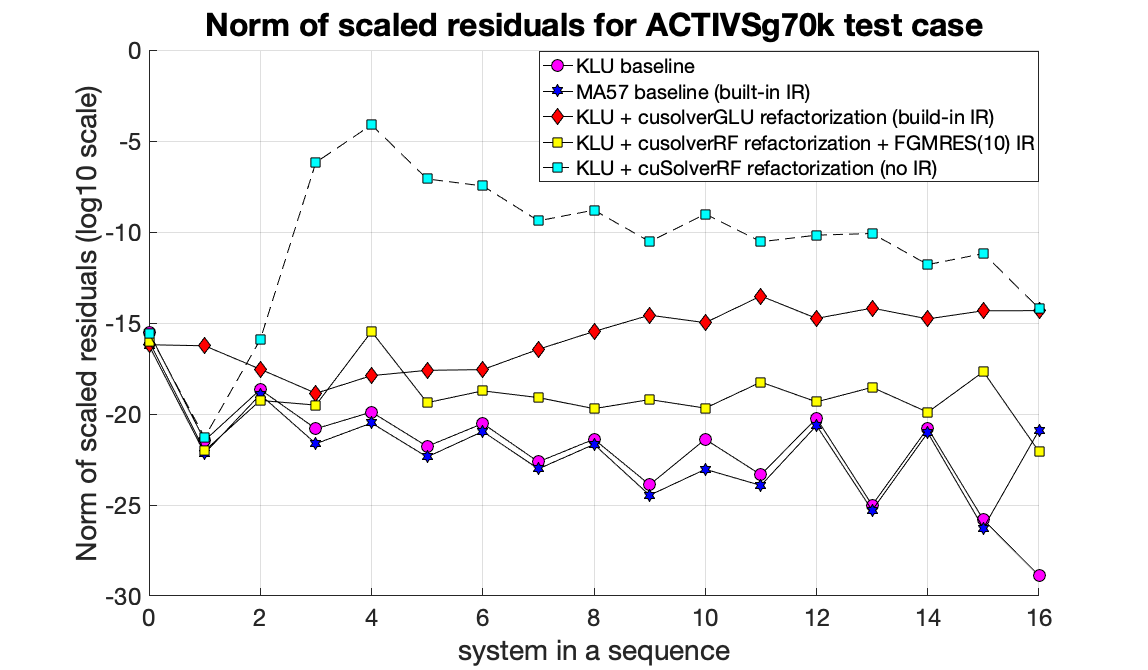}
    \caption{Norm of scaled residuals computed with different solver strategies for the largest standalone test case series. Refactorization 
    without iterative refinement (cyan squares) produces solutions of insufficient quality. The refactorization approach with our iterative refinement (yellow squares) delivers solution quality comparable to that of solvers that perform full numerical factorization for each system, and better quality than iterative refinement implemented in \cusolverglu (red diamonds). }
    \label{fig:ir_standalone_error}
\end{figure}

\paragraph{Numerical error} First, we test how effective iterative refinement is in reducing the error, based on the MA57 measure of error (\ac{nsr}). Figure~\ref{fig:ir_standalone_error} compares \ac{nsr} for the largest test case. One can observe that \ac{nsr} grows as soon as refactorization is used (starting at the third linear system in each sequence). \ac{fgmres} iterative refinement reduced the \ac{nsr} to levels that are reasonably close to the baseline (MA57 and KLU not using refactorization). On the other hand, the built-in iterative refinement in \cusolverglu is not as effective at reducing the \ac{nsr}.

\begin{figure}[htb]   
    \centering
    \includegraphics[width=\columnwidth]{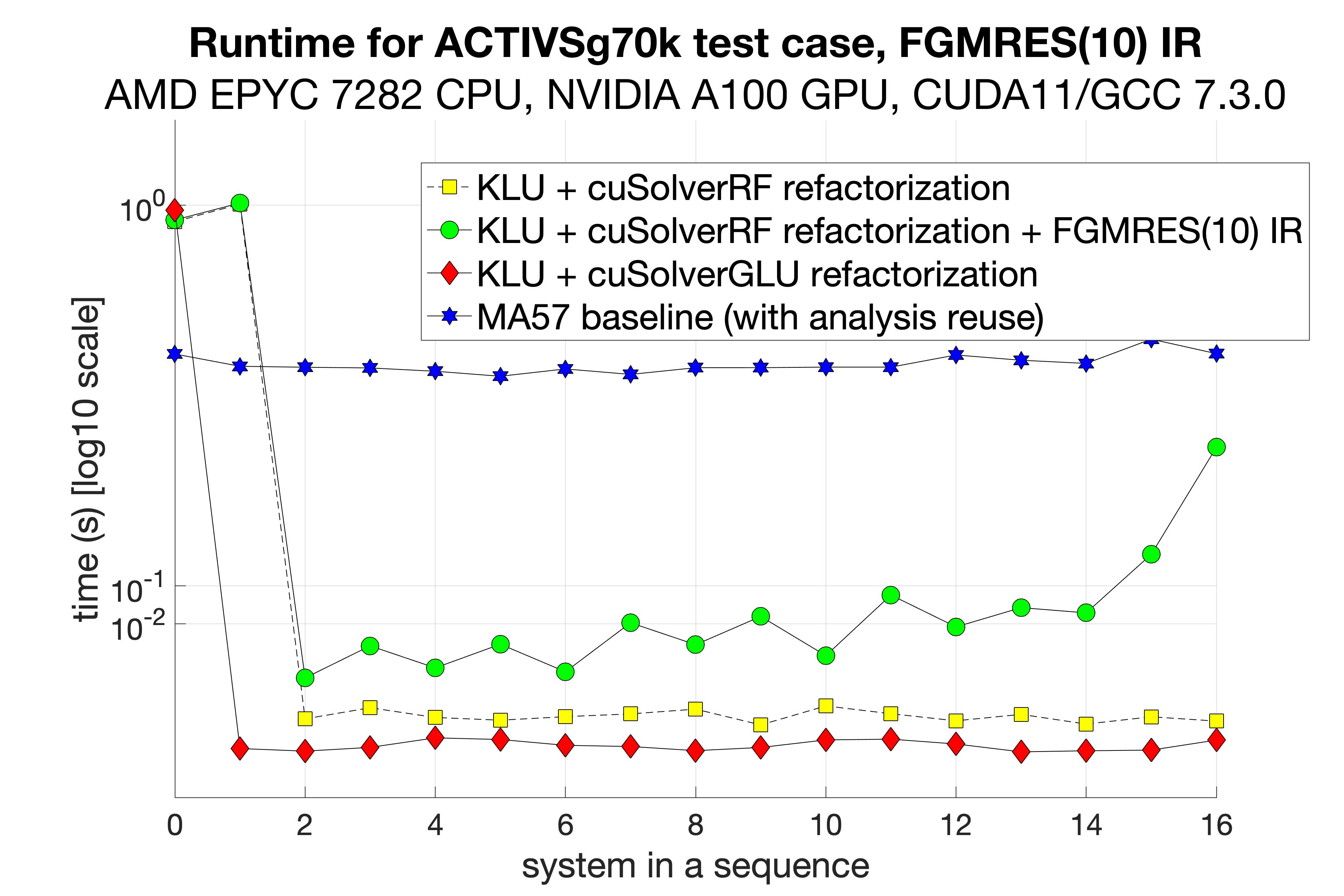}
    \caption{Timing results for the \cusolverrf with and without \ac{fgmres} iterative refinement for the largest standalone test case series (green circles and red diamonds, respectively). The tolerance for \ac{fgmres} was $10^{-14}$. The iterative refinement adds non-negligible overhead but the computation still outperforms the MA57 baseline.}
    \label{fig:ir_time_standalone}
\end{figure}

\paragraph{Computational cost} Second, Figure~\ref{fig:ir_time_standalone} compares the performance of iterative refinement for the largest test case. The results for the other two test cases are similar. As expected, iterative refinement adds a non-negligible overhead. The largest cost in the iterative refinement is the triangular solver (used as a preconditioner); see Table~\ref{tab:fgmrres_time_standalone} for details. Two conclusions can be drawn here: an iterative refinement method that requires the least number of triangular solves on the GPU is preferred, and optimizing the implementation of triangular solves will make the iterative refinement more viable. 

\begin{figure}[tb]   
    \centering
    \includegraphics[width=0.45\textwidth]{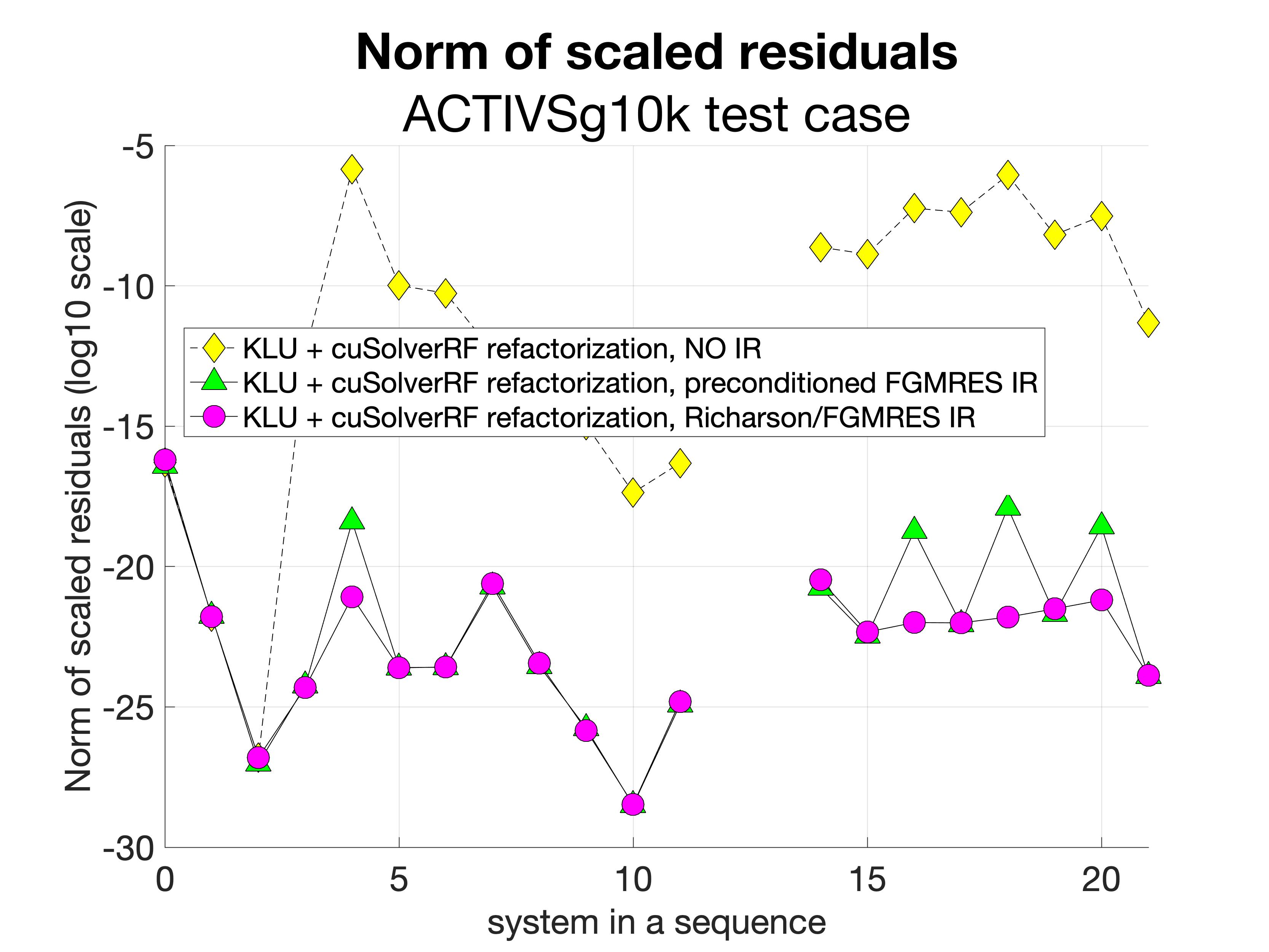}
    \includegraphics[width=0.45\textwidth]{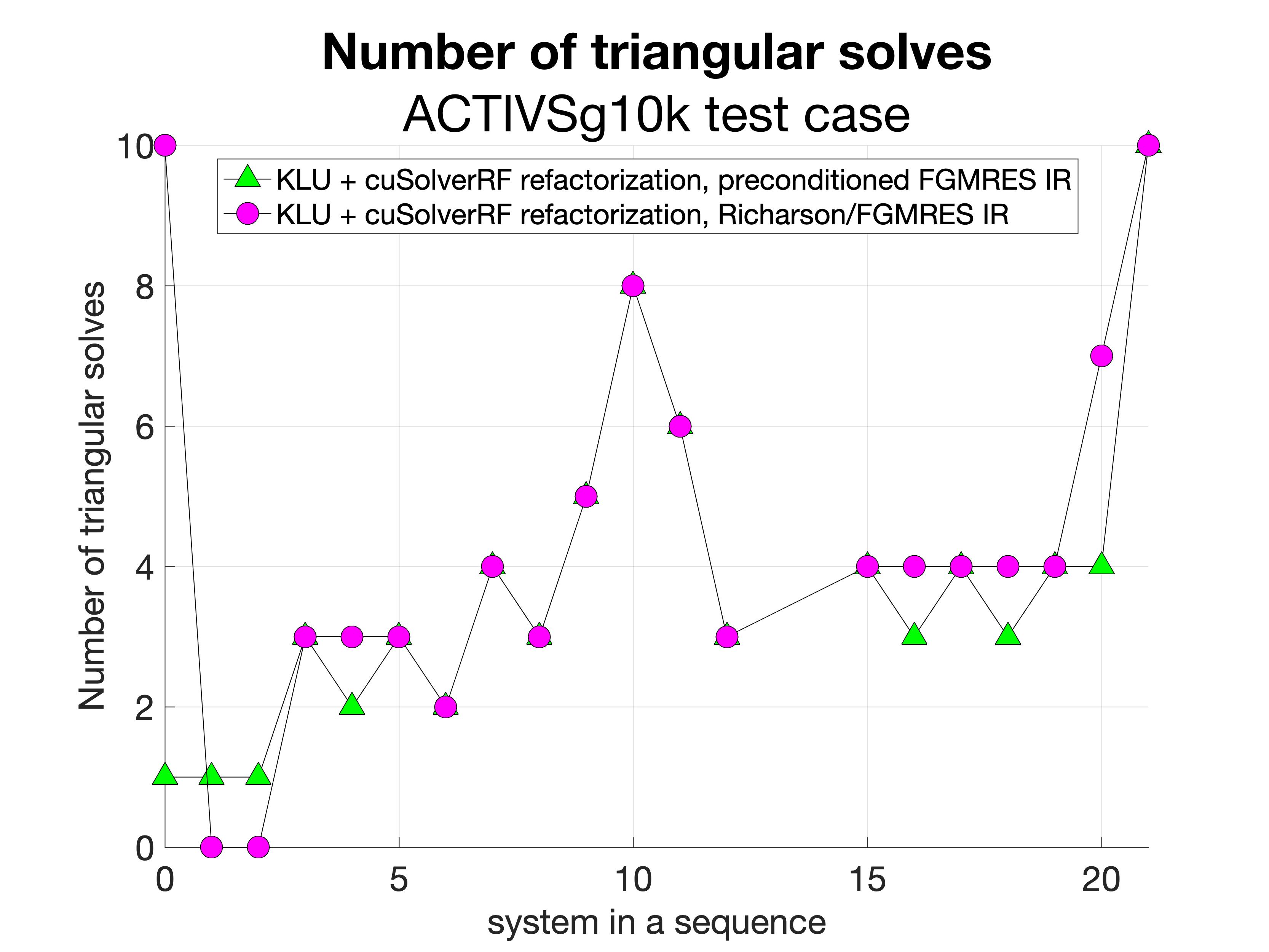}
    \caption{Our iterative refinement approach (green triangles) performs on par with Richardson-style iterations (magenta circles) and reduces norm of scaled residuals to below machine precision (left figure), while requiring fewer triangular solves overall (right). Note: outlier cases have been removed. FGMRES(20) was used in the numerical experiments.}
    \label{fig:RicharsonVsNotRicharson}
\end{figure}

\paragraph{Richardson-style approach vs iterative solver approach} 
For completeness, we also tested the standard algorithm with \ac{fgmres}(10) applied in step 5 of Algorithm 1.1 in \cite{carson2017} versus our approach based on preconditioned \ac{fgmres}(10). The results show that the number of triangular solves is slightly lower in our approach, while the residual norm reduction is comparable; see Figure~\ref{fig:RicharsonVsNotRicharson}.

\begin{table}[htb]   
    \caption{Breakdown of average cost of iterative refinement per linear system. Column IR contains average number of iterations and runtime for \ac{fgmres}(10) iterative refinement. The other two columns contain figures for average time in seconds and as percentage of the total time for triangular solve and Gram-Schmidt reorthogonalization. The iterative refinement cost is dominated by the triangular solve.}
    \label{tab:fgmrres_time_standalone}
    \centering
    \renewcommand{\arraystretch}{1.25}

    \scalebox{0.72}{
    \begin{tabular}{c|lr|lr|lr}
        \toprule
        \multirow{2}{*}{Test case} & \multicolumn{2}{c|}{IR}  & \multicolumn{2}{c|}{Triangular Solve} & \multicolumn{2}{c}{Gram-Schmidt}   \\
        \cline{2-7}
                   & iters & time (s) & time (s) & \% total & time (s) & \% total  \\
        \hline
        \hline
        ACTIVSg10k & 2.65  & 0.084    & 0.081    & 96.5 \%  & 0.003    & 3.2\%    \\
        ACTIVSg25k & 2.60  & 0.120    & 0.117    & 94.7 \%  & 0.006    & 4.9\%    \\
        ACTIVSg70k & 3.65  & 0.365    & 0.361    & 99.1 \%  & 0.002    & 0.5\%    \\
        \bottomrule
    \end{tabular}}
\end{table}


We thus decided that the linear solver prototype using \cusolverrf refactorization followed by \ac{fgmres} iterative refinement is worth interfacing to and testing with the optimization engine. Iterative refinement adds a small computational overhead, but the residual norm reduction is very promising.

\subsection{Tests in Full Software Stack}

Interfacing linear solvers with a production level application is often time-consuming and resource intensive. There are no generally accepted standards on which methods the application programming interface of linear solvers should provide. This is particularly true for direct linear solvers. Furthermore, the data formats for linear algebra objects vary among applications and linear solvers alike. This is impeding development and, consequently, impeding adoption of numerical linear algebra techniques. In what follows, we describe our \ac{acopf} application software stack, the interfacing challenges, and the results of our numerical experiments.

\subsubsection{Description of the Testbed}

Our software stack consists of the \exago power systems analysis framework \cite{ExaGo}, the \hiop  optimization library \cite{hiop_techrep}, and our linear solvers. \exago provides applications for several different analyses including multi\-period and stochastic optimal power flow. For these experiments we use \exago's \ac{acopf} module. \exago can read power grid data from common input file formats and generate power grid models including analytical expressions for Hessians and constraint Jacobians. It can use \hiop or \ipopt as an optimization engine. Both optimization packages implement an interior-point filter line-search algorithm \cite{wachter2006implementation} with some variations. 

We chose to use \hiop for our experiments because of its support for \nvidia and AMD \acp{gpu}, as well as publicly available functionality tests for convex and nonconvex optimization. \hiop also provides functionality tests for the inertia-free filter line-search algorithm \cite{chiang2016inertia}, which is required when using LU-based linear solvers. In all tests we used \hiop version 0.7.2, with slight modifications applied to its linear solver interface. While \ipopt is more mature software, it does not provide any \ac{gpu} support and has a limited number of tests available in the public domain, which makes new code verification more time-consuming.

We interfaced \hiop with our refactorization-based linear solvers using \cusolverglu (later referred to as GLU) and \cusolverrf (later referred to as Rf) refactorization modules from the \cusolver library. Our linear solvers also feature iterative refinement functions described in section \ref{sec:ir}. The criterion for deploying iterative refinement was $\|r_k-K_k\Delta x_k^{(0)}\|> \delta_\text{tol}\cdot \|r_k\|$, where $\Delta x_k^{(0)}$ was the solution given by the direct linear solver and $\delta_\text{tol}$ is a user-chosen tolerance. Here we show results for $\delta_\text{tol}=10^{-9}$ and $\delta_\text{tol}=10^{-10}$. We use standard \ac{gmres} convergence criteria based on a scaled residual norm.  Since GLU has its own built-in iterative refinement without a user option to disable it, we use our iterative refinement only with the Rf module (later referred to as Rf+IR). We use the state-of-the-art MA57 solver as our \ac{cpu} baseline.

We present results for four synthetic grid models \cite{birchfield2017tamu-cases} representative of test cases used in our investigation. Table \ref{tab:descr} provides the characteristics of the grid models evaluated. We note that the linear systems generated by \exago are slightly different from those in Table \ref{tab:standalone} as they are generated by a different tool. 

\begin{table}[htb]  
    \caption{Characteristics of the test networks specifying the number of buses, generators, and lines. The specifics of the linear system (\ref{eq:kktlinear}) for each of these networks are given in terms of the matrix size (N) and number of nonzeros (nnz) in the lower triangular part (including the diagonal) of the matrix $K_k$. }
    \label{tab:descr}
    \centering
    \scalebox{0.7}{
    \begin{tabular}{l|r|r|r|r|r} 
        \toprule
        U.S. Grid Model    & Buses & Generators & Lines    & $N(K_k)$ & nnz($K_k$) \\ 
        \hline
        \hline
        Western (simple)    & 10,000 &  2,485  &  12,706  &   52,214 &   649,168  \\
        Western             & 30,000 &  3,526  &  35,393  &  146,798 & 1,715,816  \\
        Eastern             & 70,000 & 10,390  &  88,207  &  361,134 & 4,366,256  \\ 
        Eastern + Western   & 80,000 & 12,875  & 100,923  &  413,338 & 5,015,436  \\ 
        \bottomrule
    \end{tabular}}
\end{table}

\subsubsection{Experimental results}
\label{sec:results-application}

Numerical experiments were conducted on the Summit supercomputer at the \ac{olcf}. Each node of Summit is equipped with two IBM Power 9 \acp{cpu} and six NVIDIA V100 \acp{gpu}. The codes were compiled using GCC 10.2.0 and \cuda 11.4. For each combination of grid model and choice of linear solver within the optimization engine, we solved the \ac{acopf} problem, and we attached a profiler to identify the most expensive operations and to characterize the performance of the different linear solvers and the effect of iterative refinement.

\begin{figure}[tb!]   
    \centering
    \begin{subfigure}{\columnwidth}
        {\includegraphics[width=\columnwidth,trim={0.1cm 0.1cm 0.1cm 0.1cm},clip]{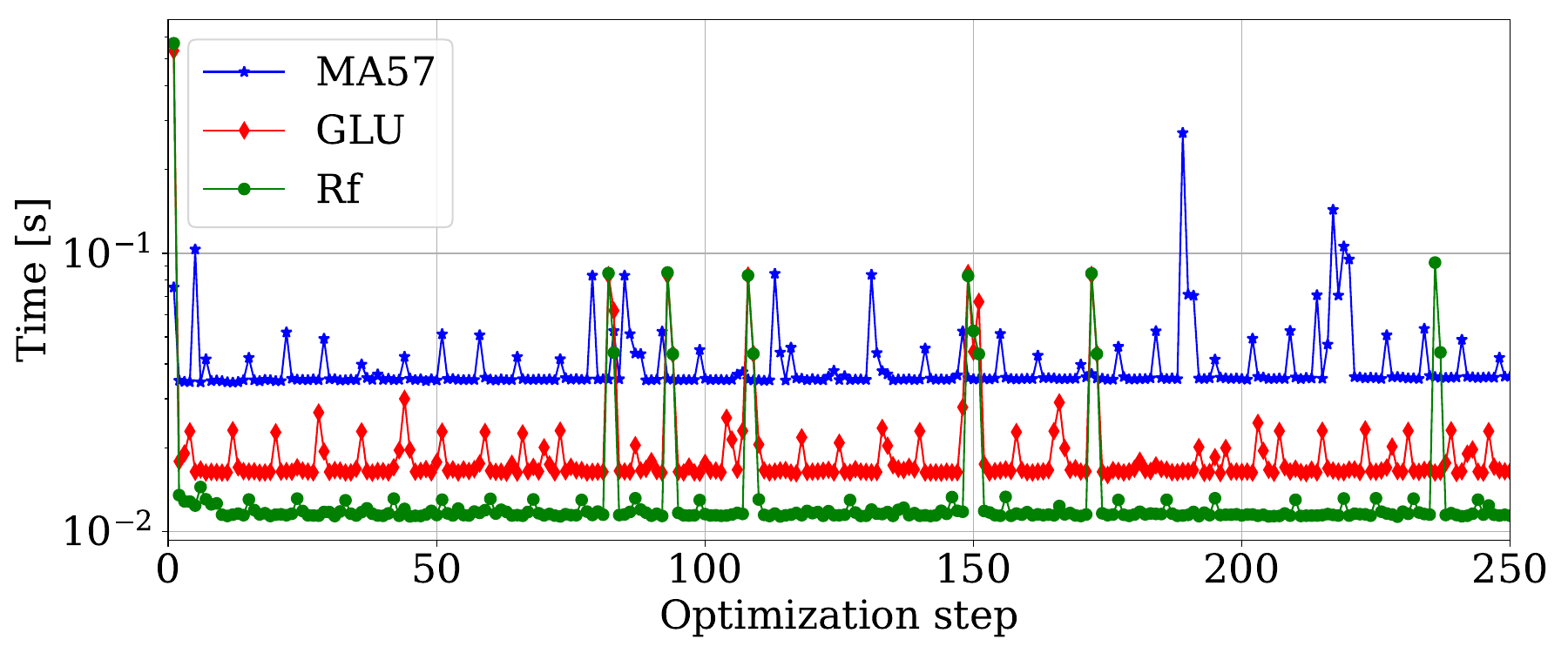}}
        \caption{Western (simple) U.S. grid}
        \label{fig:ma57_vs_cusolver.percall_a}
    \end{subfigure}
    \begin{subfigure}{\columnwidth}
        {\includegraphics[width=\columnwidth,trim={0.1cm 0.1cm 0.1cm 0.1cm},clip]{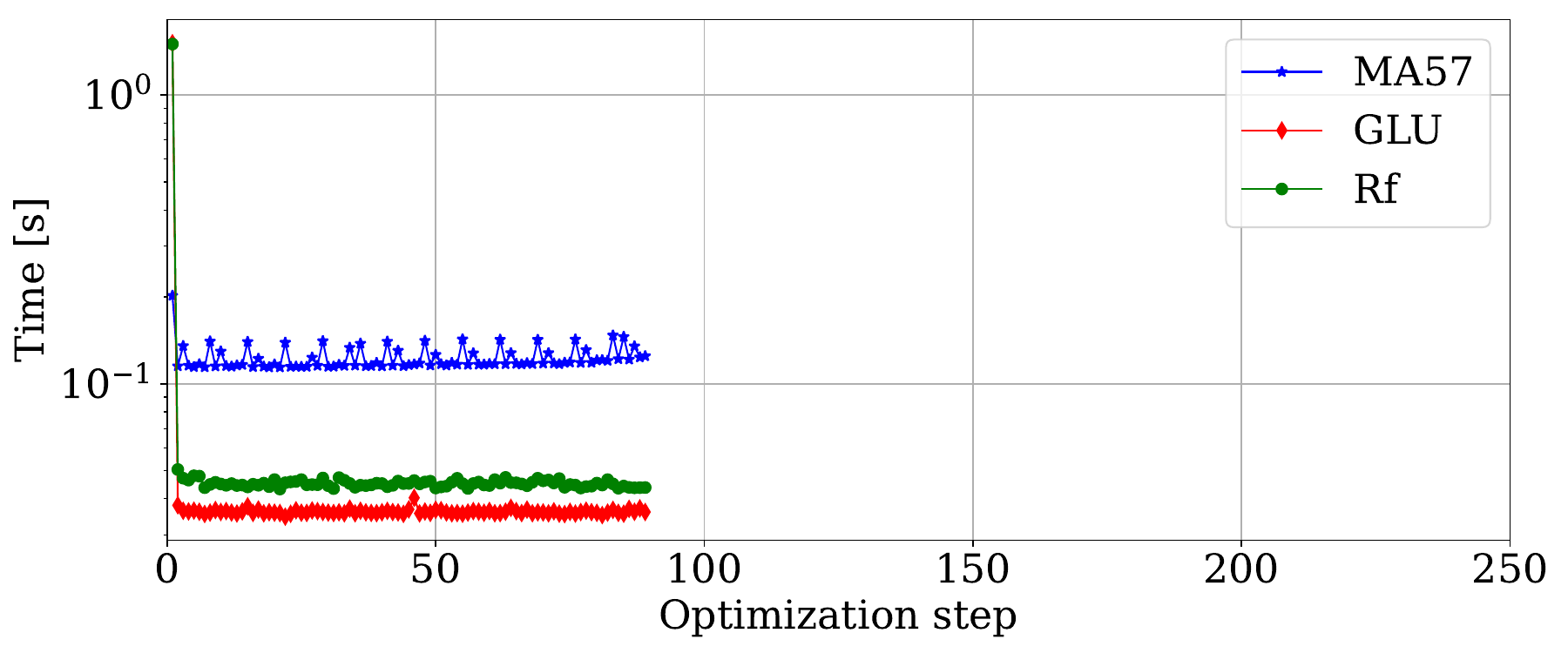}}
        \caption{Western U.S. grid}
    \end{subfigure}
    \begin{subfigure}{\columnwidth}
        {\includegraphics[width=\columnwidth,trim={0.1cm 0.1cm 0.1cm 0.1cm},clip]{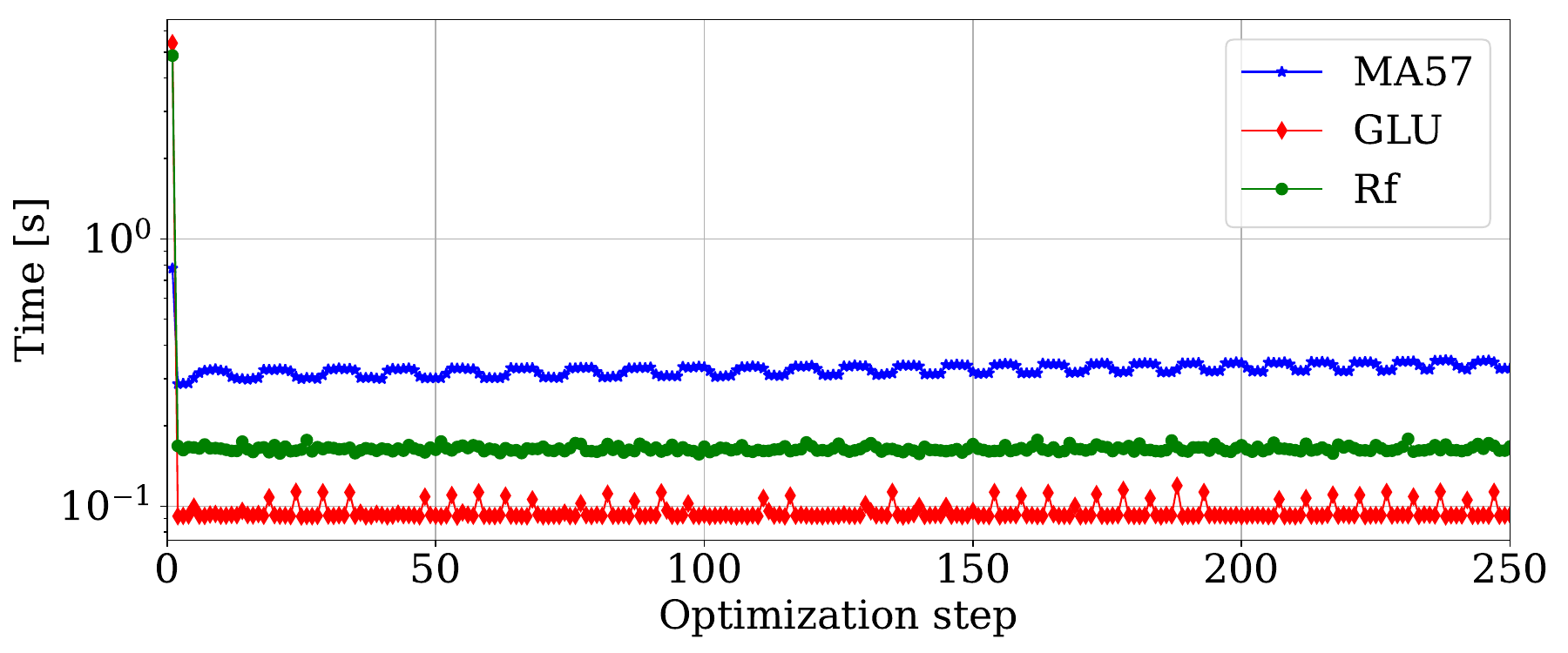}}
        \caption{Eastern U.S. grid}
    \end{subfigure}
    \begin{subfigure}{\columnwidth}
        {\includegraphics[width=\columnwidth,trim={0.1cm 0.1cm 0.1cm 0.1cm},clip]{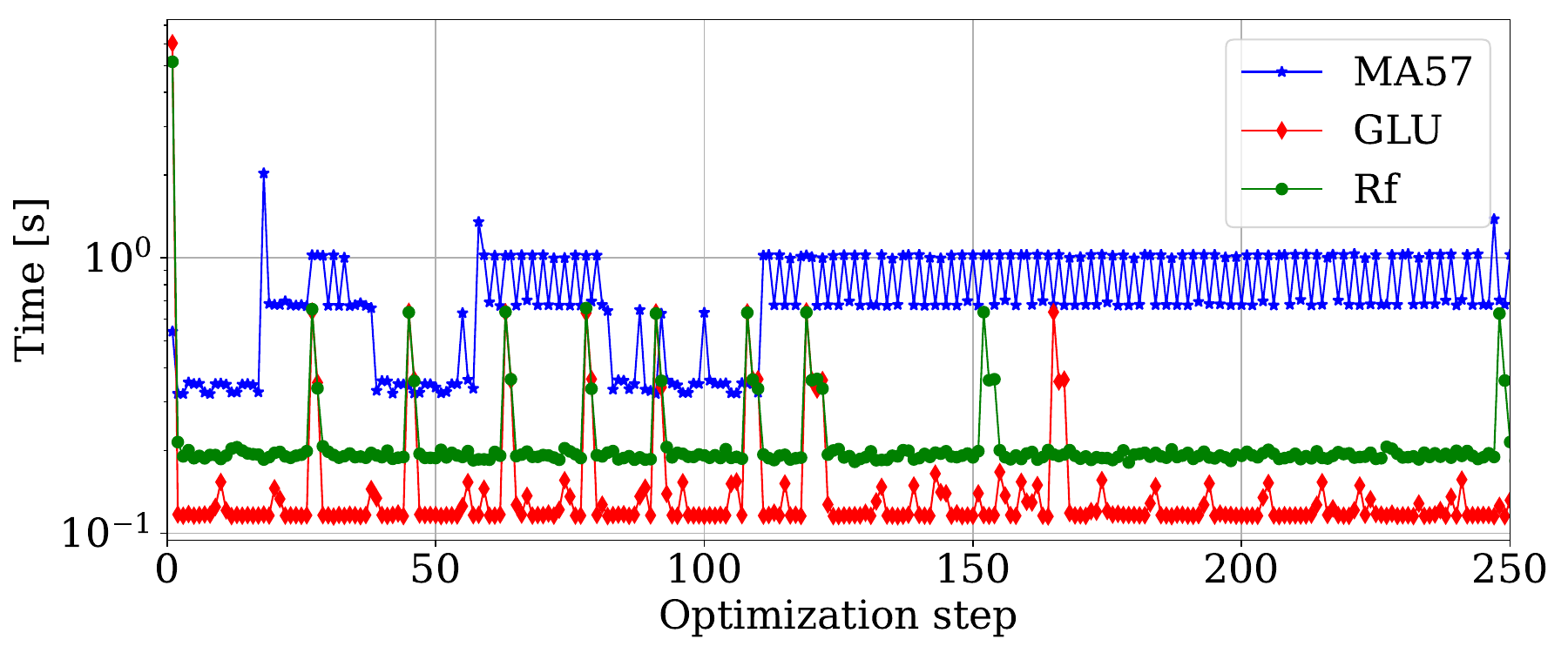}}
        \caption{Eastern+Western U.S. grid}
    \end{subfigure}
    \caption{Comparison of the cost of each matrix factorization when different linear solvers are nominally used for \ac{acopf} on \ac{olcf} Summit. Each \ac{gpu} factorization (GLU in red and Rf in green) outperforms each \ac{cpu} factorization (MA57 in blue).}
    \label{fig:ma57_vs_cusolver.percall}
\end{figure}

Figure \ref{fig:ma57_vs_cusolver.percall} provides the linear solver cost per optimization step. As expected, there is an initial cost penalty for the first factorization when \ac{gpu}-resident linear solvers are nominally employed. Recall that this first factorization is performed on the \ac{cpu} using KLU, and the data structures are transferred to the \ac{gpu}. Subsequent GLU and Rf refactorizations are typically significantly less expensive than MA57. This agrees with predictions from tests with standalone matrices (see Figure \ref{fig:ir_time_standalone}). The difference from tests on standalone matrices is that the optimization algorithm has internal logic to decide when to refactorize and perform triangular solves. For example, if the computed search direction $\Delta x_k$ is not deemed satisfactory, the system may be regularized, refactorized and solved multiple times until the optimization solver step is accepted. This process is independent of the linear solver iterative refinement. Furthermore, \hiop has a feasibility restoration mechanism that internally changes the linear solver to MA57. This occurred a few times in our tests on the Western (simple) and Eastern+Western U.S. grids. Figure \ref{fig:ma57_vs_cusolver.percall_a} illustrates this, as the otherwise lower timings of the GLU and Rf linear solvers exhibit spikes that reach the MA57 timing (e.g., call 82, 93 and 149). Therefore, the performance of the linear solvers is coupled with the actions of the optimization algorithm in the following results.

\begin{figure}[tb!] 
    \centering
    {\includegraphics[width=\columnwidth,trim={0cm 0cm 0cm 0cm},clip]{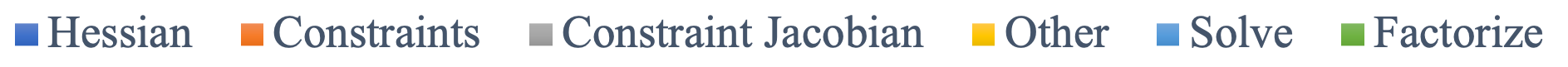}} \\
    \begin{subfigure}{\columnwidth}
        {\includegraphics[width=\columnwidth,trim={0.1cm 0.1cm 0.1cm 0.1cm},clip]{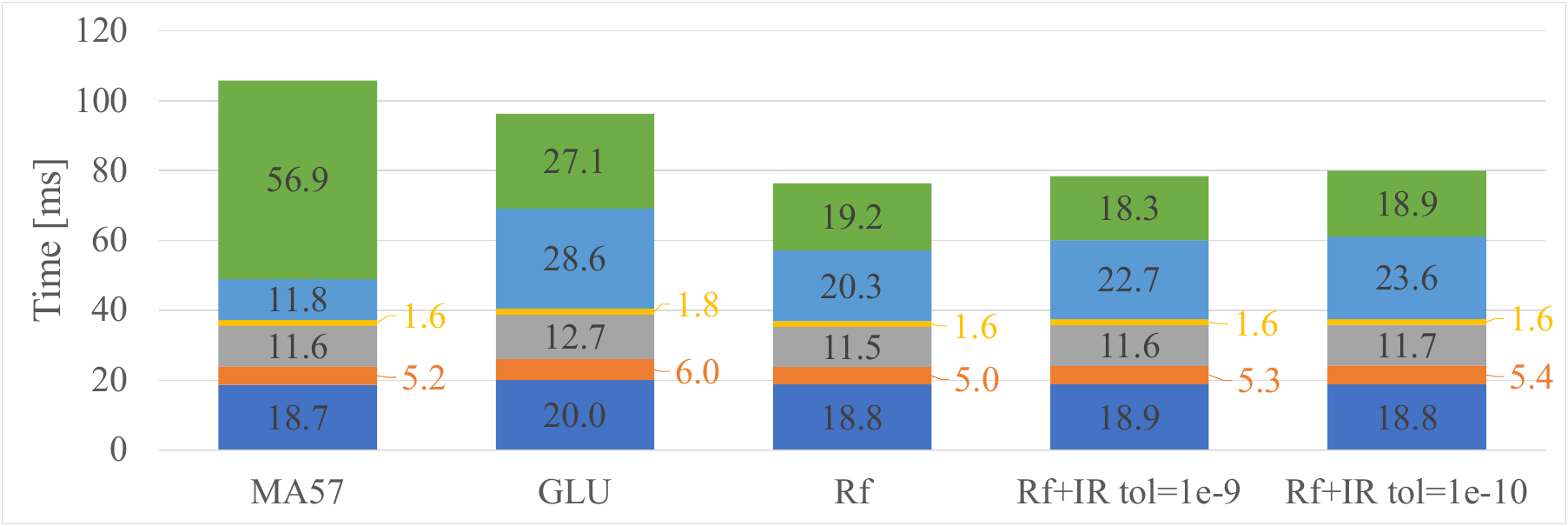}}
        \caption{Western (simple) U.S. grid}
    \end{subfigure}
    \begin{subfigure}{\columnwidth}
        {\includegraphics[width=\columnwidth,trim={0.1cm 0.1cm 0.1cm 0.1cm},clip]{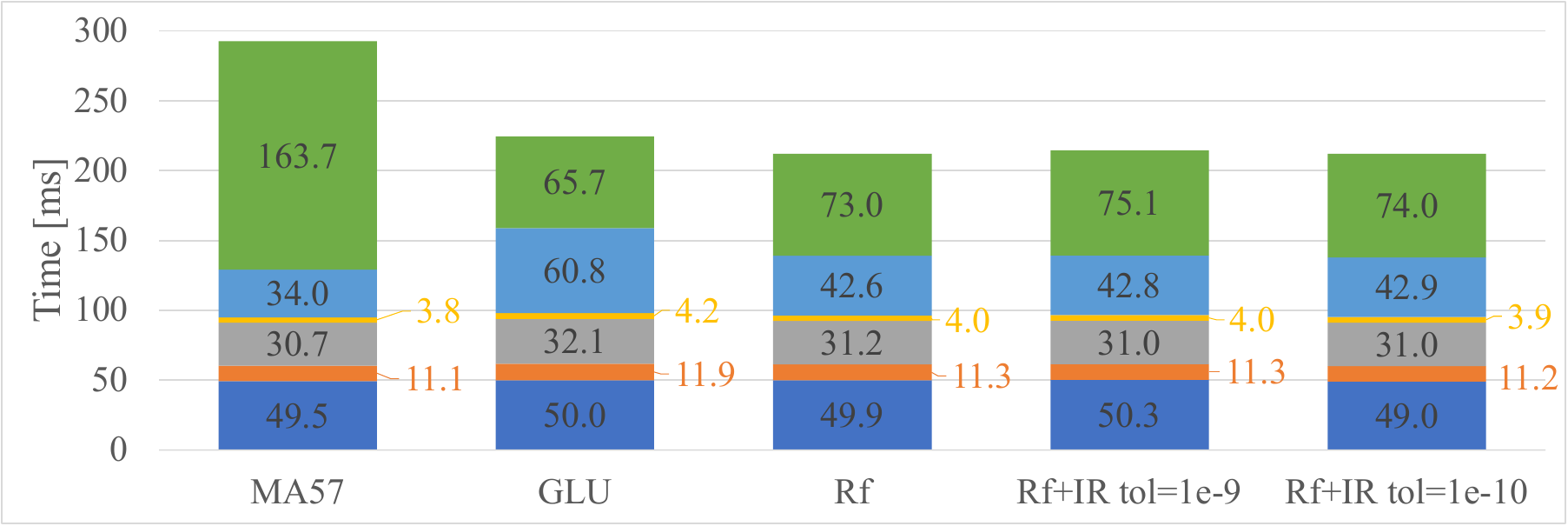}}
        \caption{Western U.S. grid}
    \end{subfigure}
    \begin{subfigure}{\columnwidth}
        {\includegraphics[width=\columnwidth,trim={0.1cm 0.1cm 0.1cm 0.1cm},clip]{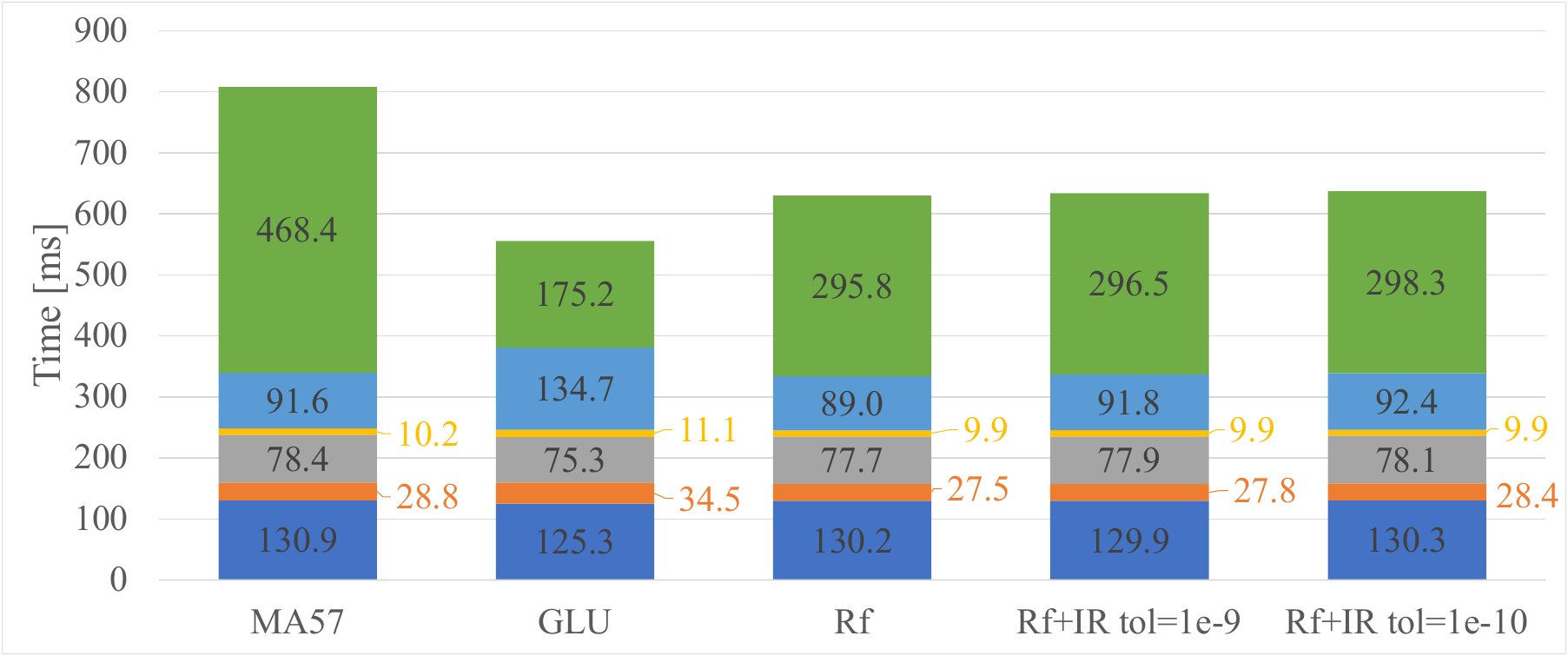}}
        \caption{Eastern U.S. grid}
    \end{subfigure}
    \begin{subfigure}{\columnwidth}
        {\includegraphics[width=\columnwidth,trim={0.1cm 0.1cm 0.1cm 0.1cm},clip]{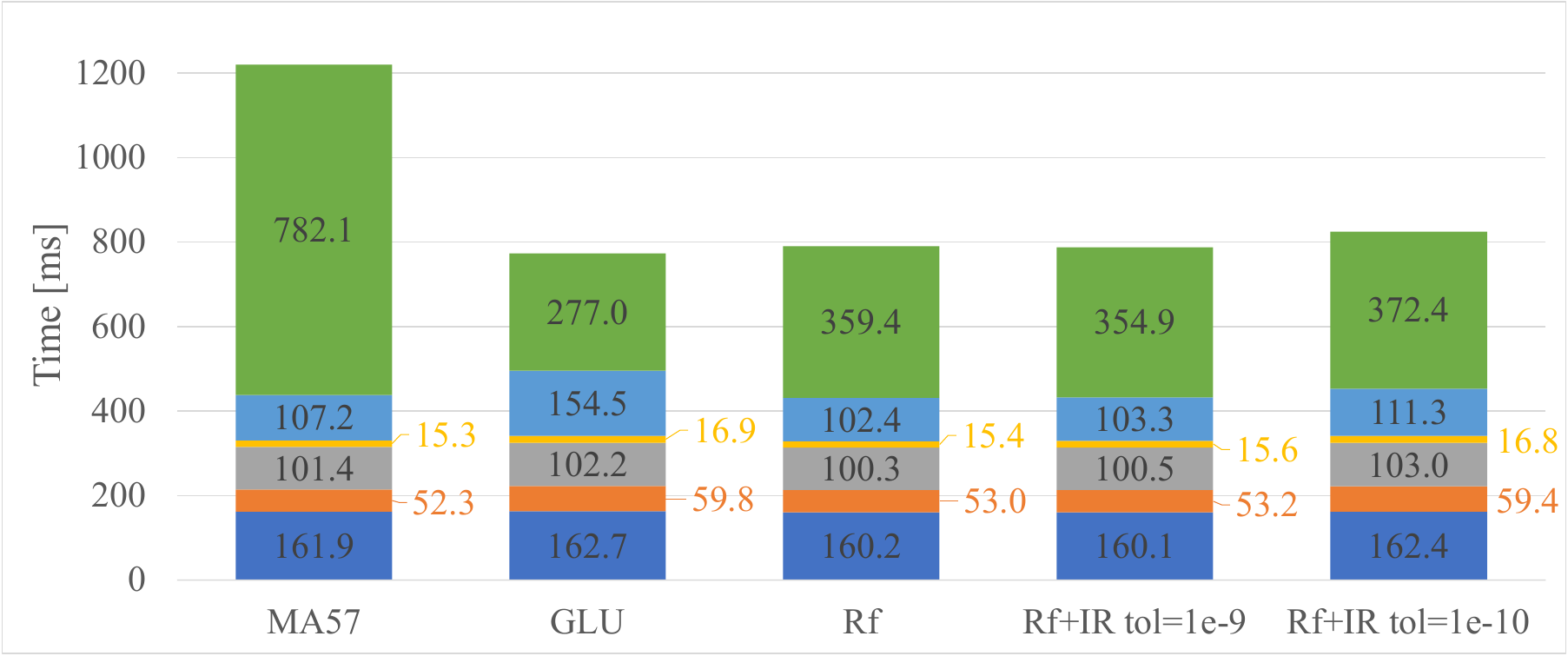}}
        \caption{Eastern+Western U.S. grid}
    \end{subfigure}
    \caption{Comparison of the average computational cost per iteration of most expensive operations when different linear solvers are used for \ac{acopf} on \ac{olcf} Summit. The cost of the first step, which is executed on \ac{cpu}, is accounted for in the averages.}
    \label{fig:ma57_vs_cusolver.averageiter}
\end{figure}

Figure \ref{fig:ma57_vs_cusolver.averageiter} provides a comparison among different linear solvers in terms of average computational cost per optimization solver step. The ``Factorize'' bars represent the average cost of matrix factorization, taking into account the (expensive) first factorization on \ac{cpu}. The ``Solve'' bars represent the cost of triangular solve and iterative refinement. All other computations including Hessian and Jacobian evaluations are performed on \ac{cpu} and their average performance per optimization step is not affected by changing the linear solver.

The \ac{gpu}-resident linear solvers outperform MA57 by $1.2\times$ to $2\times$ with the performance improvements increasing with linear system size. This suggests that the cost of the first factorization is amortized by reusing sparsity patterns of LU factors over subsequent optimization steps and that the refactorization approach scales well with linear system size. 

All acceleration on \ac{gpu}, however, comes from the matrix factorization. The solve function in GLU is actually slower than the equivalent MA57 function. More detailed profiling results indicate that this is due to redundant refinement iterations. As we mentioned earlier, GLU does not provide user options to configure or turn off the iterative refinement. The solve function in Rf performs better and is at par with its MA57 equivalent for larger test cases. Our iterative refinement implementation adds minimal overhead to the Rf solve function (Figure \ref{fig:ma57_vs_cusolver.averageiter}).

The impact of iterative refinement is in controlling the quality of the linear system solution and thus affecting the convergence of the optimization algorithm calling the linear solver. 
Table \ref{tab:ma57_vs_cusolver} is a comparison of our \ac{gpu} linear solvers evaluated with the \ac{cpu}-based MA57 in terms of the overall run statistics: the total time and the number of optimization solver steps. 

\begin{figure}[htb!]   
    \centering
    {\includegraphics[width=\columnwidth,trim={0cm 0cm 0cm 0cm},clip]{./profile_legend}} \\
    {\includegraphics[width=\columnwidth,trim={0.1cm 0.1cm 0.1cm 0.1cm},clip]{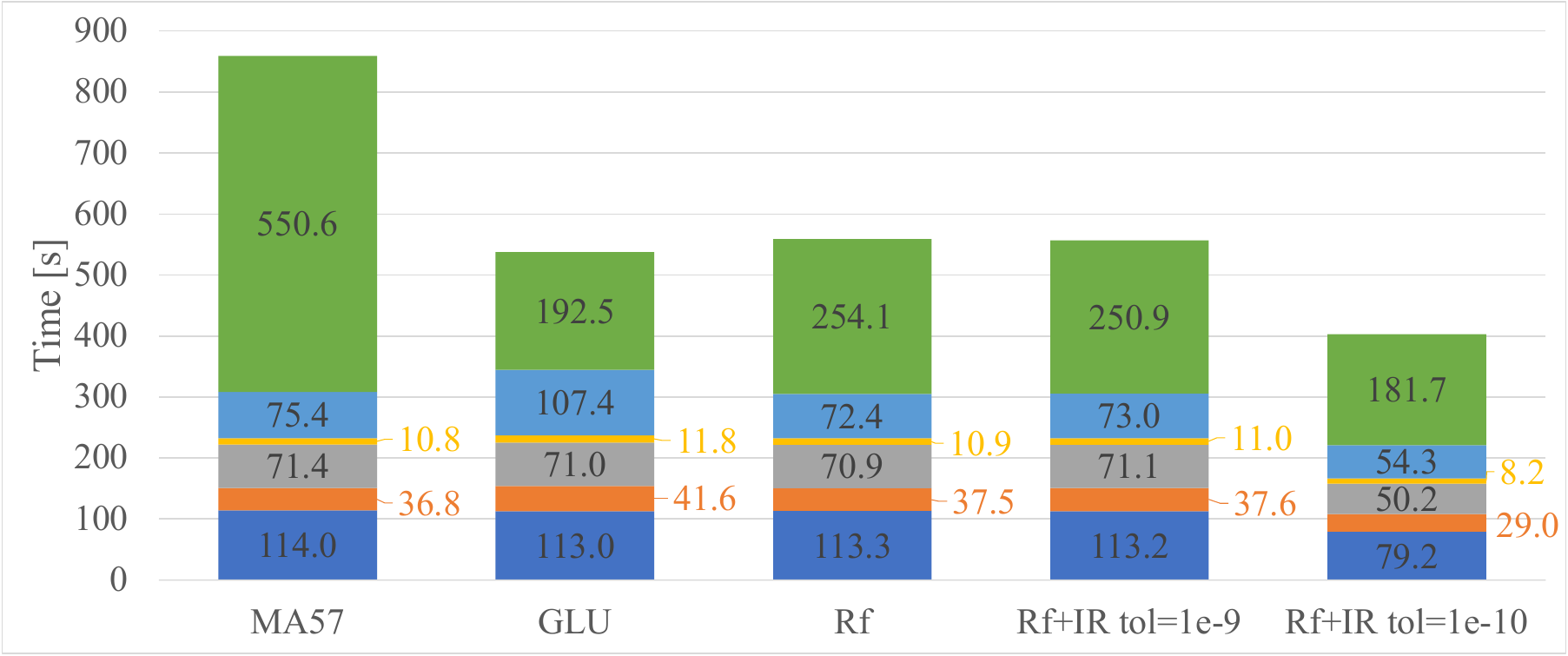}}
    \caption{Comparison of the total computational cost of most expensive operations when different linear solvers are used for \ac{acopf} on \ac{olcf} Summit. Eastern+Western U.S. grid. Time spent in factorizations is significantly reduced on \ac{gpu}, and applying iterative refinement after \cusolverrf with $\delta_\text{tol}=10^{-10}$ results in the smallest number of optimization solver steps and total runtime.}
    \label{fig:ma57_vs_cusolver}
\end{figure}

We see that in all cases, iterative refinement reduces the number of optimization solver steps to solution except in the case of Western U.S. grid model where it was not needed (refactorization produced sufficiently accurate solutions). In the case of the Eastern + Western U.S. grid model, iterative refinement with a tolerance of $10^{-10}$ significantly reduces the number of optimization solver steps, as shown in the last column of Table \ref{tab:ma57_vs_cusolver}, resulting in a shorter time to the converged solution. More specifically, this configuration of iterative refinement (Rf+IR with tolerance $10^{-10}$) results in $40\%$ speedup of the overall \ac{acopf} analysis over the configuration with Rf linear solver but without iterative refinement, $30\%$ speedup over the configuration with GLU, and $100\%$ speedup over the MA57 baseline linear solver (see Figure \ref{fig:ma57_vs_cusolver}). 

\begin{table*}[bt]   
    \centering
    \caption{Total run times for \ac{acopf} when using MA57, GLU and Rf linear solvers on \ac{olcf} Summit. The number of steps is the total number of optimization solver steps to the converged solution.}
    \scalebox{0.85}{
    \begin{tabular}{m{0.35\columnwidth}|>{\centering\arraybackslash}m{0.07\columnwidth}|>{\centering\arraybackslash}m{0.1\columnwidth}|>{\centering\arraybackslash}m{0.07\columnwidth}|>{\centering\arraybackslash}m{0.1\columnwidth}|>{\centering\arraybackslash}m{0.07\columnwidth}|>{\centering\arraybackslash}m{0.1\columnwidth}|>{\centering\arraybackslash}m{0.07\columnwidth}|>{\centering\arraybackslash}m{0.1\columnwidth}}
        \toprule
        \multirow{3}{*}{Solver} & \multicolumn{8}{c}{U.S. Grid Model} \\ 
        \cline{2-9}
        & \multicolumn{2}{c|}{Western (simple)} & \multicolumn{2}{c|}{Western} & \multicolumn{2}{c|}{Eastern} & \multicolumn{2}{c}{Eastern+Western} \\ 
        \cline{2-9}
        & t [s] & \# steps & t [s] & \# steps & t [s] & \# steps & t [s] & \# steps \\
        \hline 
        \hline
        MA57               & 42.6 & 403 & 27.2 & 93 & 212 & 262 & 853 & 704 \\ 
        \hline 
        GLU                & 36.7 & 381 & 20.9 & 93 & 146 & 262 & 537 & 695 \\ 
        \hline 
        Rf                 & 32.0 & 419 & 19.7 & 93 & 167 & 265 & 559 & 707 \\ 
        \hline 
        {Rf+IR $\delta_\text{tol}=10^{-9}$}  
                           & 29.2 & 373 & 19.9 & 93 & 166 & 262 & 557 & 707 \\ 
        \hline 
        {Rf+IR $\delta_\text{tol}=10^{-10}$}  
                           & 30.0 & 375 & 19.7 & 93 & 167 & 262 & 403 & 488 \\ 
        \bottomrule
    \end{tabular}}
    \label{tab:ma57_vs_cusolver}
\end{table*}

\section{Specialized Linear Solvers: \hykkt}
\label{sec:hykkt}

Most interior method implementations transform 
\eqref{eq:kktnonsymmetric} into a smaller symmetric system (\ref{eq:kktlinear}), which can be solved more efficiently. The \hykkt linear solver~\cite{shaked2022,regev2022kkt,regev2022github} extends this idea and performs a more aggressive transformation to obtain a linear system form that is more suitable for solving on heterogeneous architectures. Following \cite{GG2003} it further modifies \eqref{eq:kktlinear} to
\begin{align} \label{eq:kktlineargamma}
\overbrace{\begin{bmatrix}
              H + D_x + \gamma J^TJ & J^T
           \\ J   & 0 
           \end{bmatrix}}^{K_k}
  \overbrace{\begin{bmatrix}
    \Delta x \\ 
    \Delta \lambda
  \end{bmatrix}}^{\Delta x_k}=
  \overbrace{\begin{bmatrix}
    r_{x} + \gamma J^Tr_{\lambda}\\ r_{\lambda}
  \end{bmatrix}}^{r_k},
\end{align}
by adding $\gamma J^T$ times the second row to the first row (this does not change the solution to the system). The authors of the method showed that if $J$ has full rank, $H_\gamma\doteq H + D_x + \gamma J^TJ$ is guaranteed to be symmetric positive definite ({SPD}) for large enough $\gamma$~\cite{regev2022kkt}. This system transformation allows \hykkt to solve symmetric indefinite \ac{kkt} systems by solving SPD systems. However, \hykkt is specialized for solving systems with \ac{kkt} structure as opposed to the \LDLT approach, which allows solving \textit{any} symmetric (possibly indefinite) system. 

\hykkt solves \eqref{eq:kktlineargamma}  by creating the SPD Schur-complement $S=JH_\gamma^{-1}J^T$ operator. Note that $S$ is dense, so it is not formed explicitly; it is applied as a sequence of matrix-vector products within the conjugate gradient (CG) method~\cite{hestenes1952methods}. The inverse of 
$H_\gamma$ is applied as triangular solves with its Cholesky factors. The nonzero structure of $K_k$ in \eqref{eq:kktlineargamma} is unchanged between iterations, and this property is leveraged to compute the symbolic factorization of $H_\gamma$ only once and reuse it for all subsequent $k$. Unlike \LDLT, Cholesky factorization does not require numerical pivoting, so the emphasis can be put on reducing fill-in, and no memory allocation is needed beyond the first iteration. This allows for efficient GPU utilization. 

We analyze the run-time performance of \hykkt on the linear systems from Table~\ref{tab:standalone}. All testing in this section is done using a prototype C++/\cuda code running on a single GPU device. We used a system with AMD EPYC 7742 \acp{cpu} and NVIDIA Ampere 100 \acp{gpu}. The codes were compiled using GCC 9.4.0 and \cuda 11.8.

\begin{figure}[tb!]   
    \centering
    \begin{subfigure}[t]{.49\linewidth}
        \centering\includegraphics[width=\columnwidth]{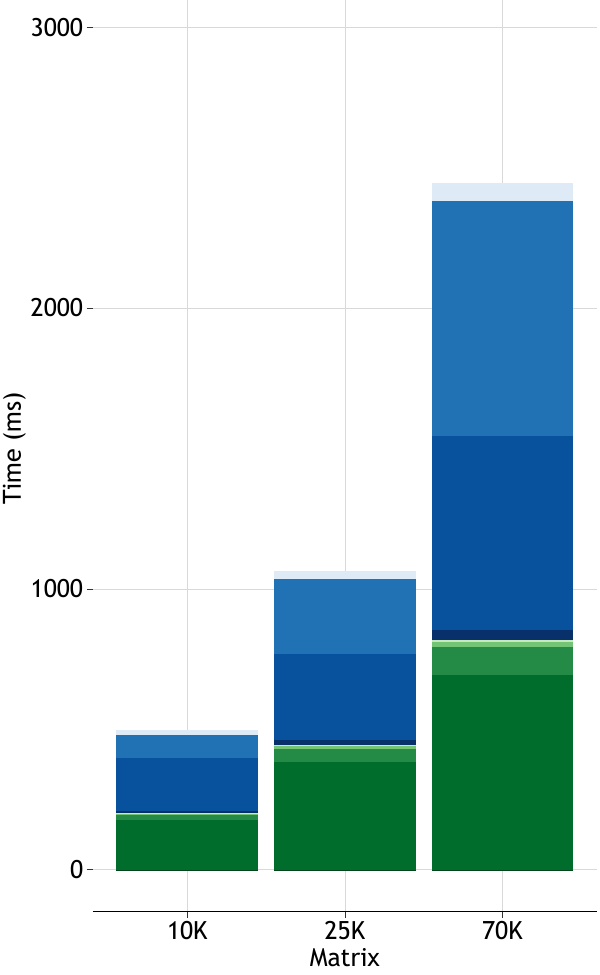}
        \caption{Initial Sequence}\label{fig:hykkt_timings_a}
    \end{subfigure}
    \begin{subfigure}[t]{.49\linewidth}
        \centering\includegraphics[width=\columnwidth]{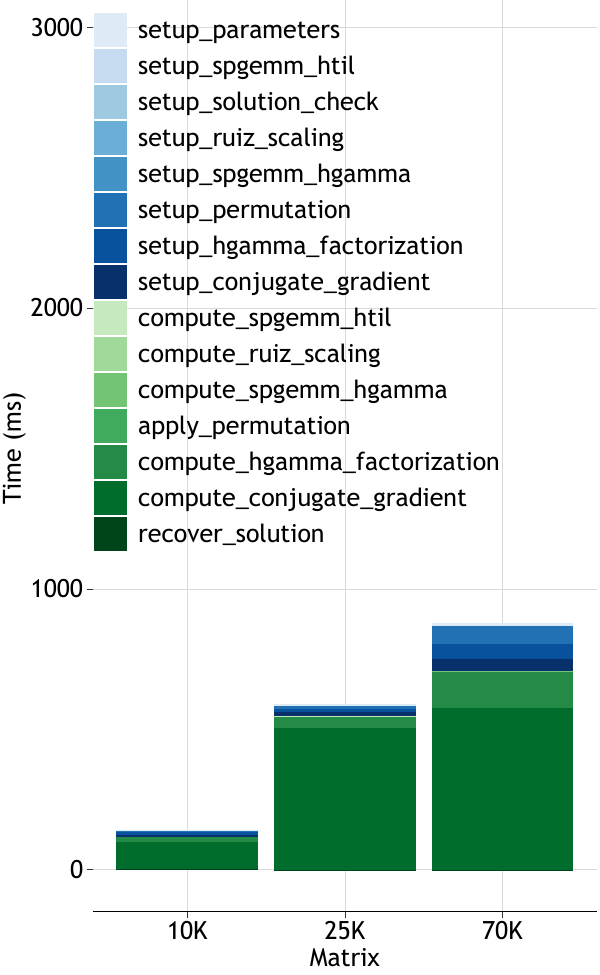}
        \caption{Average of Sequences with Reuse}\label{fig:hykkt_timings_b}
    \end{subfigure}
    \caption{Timing distribution \hykkt by solution steps.}\label{fig:hykkt_timings}
\end{figure}

In Figure~\ref{fig:hykkt_timings}, we visually depict a breakdown of timings of different stages of the solution process in \hykkt for different test cases. In these test cases, we setup 
\hykkt with an initial matrix and use the data structures created for the first system to solve a series of systems with similar sparsity patterns.  

\begin{figure}[tb!] 
    \centering
    \begin{subfigure}{\columnwidth}
        {\includegraphics[width=\columnwidth]{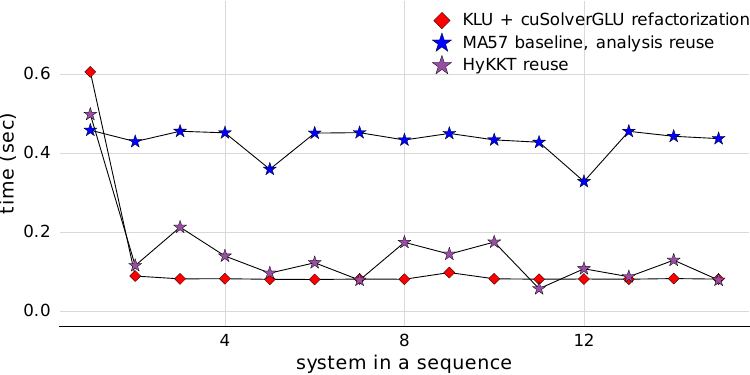}}
        \caption{ACTIVSg10k grid model}
        \vspace{5pt}
    \end{subfigure}
    \begin{subfigure}{\columnwidth}
        {\includegraphics[width=\columnwidth]{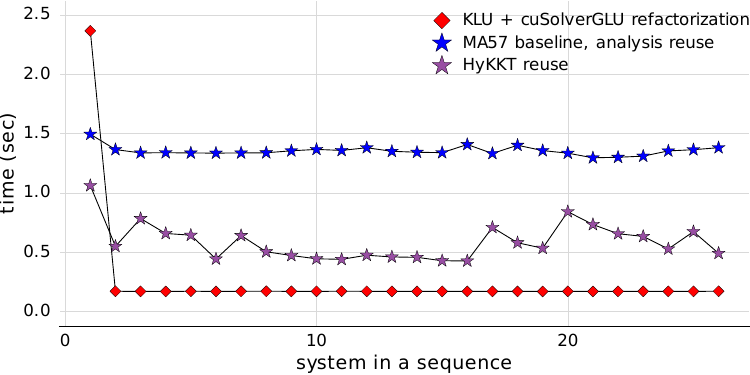}}
        \caption{ACTIVSg25k grid model}
        \vspace{5pt}
    \end{subfigure}
    \begin{subfigure}{\columnwidth}
        {\includegraphics[width=\columnwidth]{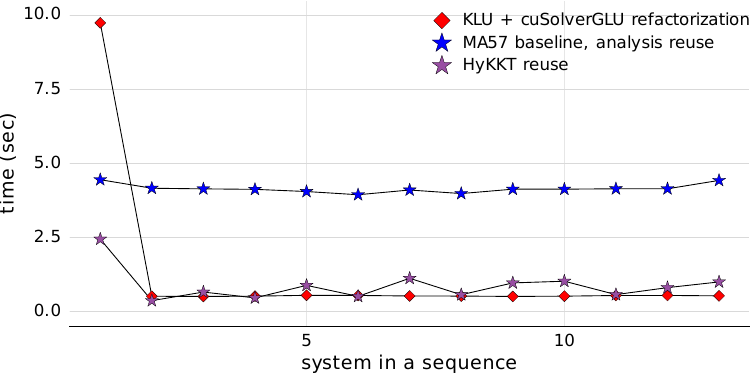}}
        \caption{ACTIVSg70k grid model}
        \label{fig:activsg_compare_c}
    \end{subfigure}
    \caption{Performance comparison for different linear solvers using ACTIVSg standalone test cases (Table \ref{tab:standalone}). HyKKT performance on \ac{gpu} is on par with \cusolverglu, but its setup cost is up to $4\times$ lower than for KLU with the \cusolverglu refactorization scheme. Both outperform the MA57 baseline.}
    \label{fig:activsg_compare}
\end{figure}

Figure~\ref{fig:hykkt_timings_a} presents the computation time for different stages of the initial linear system solve, and Figure~\ref{fig:hykkt_timings_b} shows the average computation time of each step for the remaining sequences. As observed from Figure~\ref{fig:hykkt_timings_a}, setup times dominate the cost of the computation for the first step in the sequence. As the setup is only done once, its cost is amortized over the number of systems solved. The solve time for the entire series is dominated by CG. This suggests \hykkt will scale well with the linear system size. In Table~\ref{tab:hykkt_stepwise_timing}, we present the detailed timings of different stages used in \hykkt during the initial solve. The setup functions happen only once for the entire optimization problem, and for the largest problem they are roughly $2/3$ of the cost of the first iteration. This means that the cost for solving the entire optimization problem drops by roughly a factor of $3$, as verified in Figure~\ref{fig:activsg_compare}. Further details on these functions are in the \hykkt repository \cite{regev2022github}.

\begin{table*}[tb!] 
\centering
\caption{Breakdown of \hykkt timings by the solution steps. Symbolic analysis and Cholesky factorization are the most expensive parts of the computation but need to be performed only once, for the first system in the series. CG is the most expensive operation that needs to be performed for each system in the series; it scales well with problem size.}
\label{tab:hykkt_stepwise_timing}

\scalebox{0.8}{
\begin{tabular}{@{}lrrrrrr@{}}   
    \toprule
    \multirow{2}{*}{Function Name} & \multicolumn{2}{c}{10K}     & \multicolumn{2}{c}{25K}         & \multicolumn{2}{c}{70K}\\ 
    \cmidrule(l){2-7} 
                               & Time [ms]     & \%              & Time [ms]     & \%              & Time [ms]     & \%     \\
    \midrule
setup\_parameters              & 19.140        & 3.84            & 26.647        & 2.51            & 62.804        & 2.57   \\
setup\_spgemm\_htil            & 0.003         & 0.00            & 0.009         & 0.00            & 0.009         & 0.00   \\
setup\_solution\_check         & 0.606         & 0.12            & 0.875         & 0.08            & 0.966         & 0.04   \\
setup\_ruiz\_scaling           & 0.471         & 0.09            & 0.432         & 0.04            & 0.464         & 0.02   \\
setup\_spgemm\_hgamma          & 0.001         & 0.00            & 0.001         & 0.00            & 0.001         & 0.00   \\
setup\_permutation             & 82.324        & 16.51           & 266.561       & 25.10           & 834.809       & 34.15  \\
setup\_hgamma\_factorization   & 186.539       & 37.41           & 304.347       & 28.66           & 694.077       & 28.39  \\
setup\_conjugate\_gradient     & 9.661         & 1.94            & 19.757        & 1.86            & 34.847        & 1.43   \\
\midrule[0.1pt]
compute\_spgemm\_htil          & 1.755         & 0.35            & 3.395         & 0.32            & 5.844         & 0.24   \\
compute\_ruiz\_scaling         & 0.018         & 0.00            & 0.016         & 0.00            & 0.018         & 0.00   \\
compute\_spgemm\_hgamma        & 4.062         & 0.82            & 11.036        & 1.04            & 16.790        & 0.69   \\
apply\_permutation             & 0.026         & 0.01            & 0.025         & 0.00            & 0.032         & 0.00   \\
compute\_hgamma\_factorization & 19.048        & 3.82            & 44.530        & 4.19            & 99.763        & 4.08   \\
compute\_conjugate\_gradient   & 174.869       & 35.07           & 384.253       & 36.18           & 694.113       & 28.39  \\
recover\_solution              & 0.096         & 0.02            & 0.092         & 0.01            & 0.127         & 0.01   \\ 
    \bottomrule
\end{tabular}}
\end{table*}

Figure~\ref{fig:activsg_compare} shows run-time comparisons of the various linear solvers for the ACTIVSg10k, ACTIVSg25k and ACTIVSg70k test systems respectively. Note that the total time may not match the profiling results because of the profiler overhead. These results show that \hykkt outperforms the \ac{cpu} baseline and performs well compared to other \ac{gpu}-resident linear solvers. 

Considering that \hykkt does not have obvious parallelization bottlenecks and its setup cost is much lower than for refactorization linear solvers, it remains to be seen whether on even larger problems \hykkt starts outperforming KLU+\cusolverrf and KLU+\cusolverglu or if they stay comparable as in Figure~\ref{fig:activsg_compare_c}. 
A further advantage of \hykkt is that there is still room for implementation improvement, especially for CG (the most expensive computational step). 
Additionally, the internal CG tolerance could be managed by the optimization algorithm to reduce the computation time if a larger tolerance is sufficient for the optimization solver. 
Finally, a more accurate timing comparison would be done within the context of an optimization solver, where methods other than \hykkt require forming $K_k$ explicitly at each iteration using the calculated Jacobian and Hessian, whereas \hykkt does not. If this time is substantial (which it may be as it requires a lot of data movement at each iteration), \hykkt already has an advantage over these other methods on large problems.

\section{Conclusions and future work}
\label{sec:conclusion}

We have demonstrated efficient solution of a sequence of ill-conditioned linear systems resulting from a constrained nonlinear optimization. To outperform the state-of-the-art sparse direct linear solver MA57 while taking advantage of new heterogeneous hardware, we needed to combine multiple strategies into efficient code. To reduce the cost of factorization, we used refactorization, which also allowed us to reuse the existing data structures. As our template pattern for the factors, we used the KLU linear solver, which produced the sparsest factors. To mitigate the possible loss of quality in the solution, we developed an iterative refinement strategy based on \ac{fgmres}. We also demonstrated that with sound iterative refinement, the static pivoting approach can be effectively used in realistic large-scale computations. Further, we analyzed the performance of \hykkt, another linear solver designed for deployment on heterogeneous hardware. \hykkt implements a hybrid (iterative and direct) approach and outperforms the MA57 baseline. The results were particularly good for the largest problems. Overall, the best approaches were combinations of various algorithmic and implementation-based improvements. 

It is a common misconception that any mathematical algorithm can be accelerated by implementing it on a \ac{gpu}. In our experience, one typically needs to co-design the mathematical algorithm and its implementation in order to take advantage of heterogeneous hardware. 
In our case (using a sparse direct linear solver to solve a series of very ill-conditioned linear systems), the winning strategy is to use both a \ac{gpu} and a \ac{cpu}. However, we do not simply \emph{offload} the computationally intensive parts of the computation to the \ac{gpu}, as this would result in substantial memory traffic and possible data re-allocation. Instead, we restructured our algorithms so that all operations unsuitable for \acp{gpu} (e.g., those requiring conditional branching) are done once on \ac{cpu} and their results reused for the rest of the computation. Similarly, in our implementation we allocate data structures and workspace memory only once and reuse them for the rest of the computation, thereby minimizing memory management, which is very expensive on heterogeneous hardware. It was not only the combination of mathematical methods that helped performance, but also a matching implementation strategy that judiciously uses hardware resources.

Our approach was implemented using \nvidia libraries. Some functions we used are not yet available in AMD libraries; hence our approach is not fully portable. On the other hand, we noticed that the triangular solve used by \cusolverrf is not as efficient as the one developed in AMD's ROCm library. Combining refactorization, efficient (AMD or other) triangular solve, and our iterative refinement will probably bring additional speedups and make the code portable to multiple hardware platforms. 

Results in Figure~\ref{fig:ma57_vs_cusolver.percall} show that effects of iterative refinement are perturbed by optimization solver controls such as matrix regularization. It is sometimes difficult to isolate the effects of iterative refinement providing a higher accuracy solution, if a matrix regularization is triggered before that by the optimization solver. Future research in this area requires codesign of optimization and iterative refinement algorithm controls.
While this is beyond the scope of this work, we plan to test the developed approaches with different optimization solvers (e.g., \ipopt, in addition to \hiop). 

We first tested the presented strategies on a series of linear systems generated in a single \ac{acopf} analysis. As it turned out, the resulting systems were a good proxy for the production run (full \ac{acopf}) code. The test matrices are publicly available \cite{maack2020matrices}; hence the results can be easily reproduced and possible improvement can be demonstrated on exactly the same set of tests. Many \emph{benchmark} matrices exist, but it is rare that the matrices are available as a sequence of problems and not just one. In our case, using the properties of the sequence (non-changing sparsity pattern) helped predict linear solver performance better. We emphasize the value of standalone testing: it allows fast testing of new algorithms and approaches, and helps isolate and target critical parts of the code. We also believe that such practice can speed up development of numerical linear algebra libraries, because it gives valuable predictions of a library's performance in a production setting without resource-intensive interfacing of that library with the rest of the application software.

\section*{Acknowledgments}
\noindent
This research has been supported in part by UT-Battelle, LLC, and used resources of the Oak Ridge Leadership Computing Facility under contract DE-AC05-00OR22725 with the U.S. Department of Energy (DOE). This research was also supported by the Exascale Computing Project (17-SC-20-SC), a collaborative effort of the DOE Office of Science and the National Nuclear Security Administration. 

The authors thank Cosmin Petra and Nai-Yuan Chiang of Lawrence Livermore National Laboratory for their guidance when using the \hiop optimization solver, and Shri Abhyankar of Pacific Northwest National Laboratory for his help with using \exago software.
The authors would also like to thank LungSheng Chien and Doris Pan of \nvidia for their help with using the undocumented \cusolverglu module in the cuSOLVER library. Warm thanks also go to Phil Roth of Oak Ridge National Laboratory and Christopher Oehmen of Pacific Northwest National Laboratory for their support of this work. 

\appendix

\bibliographystyle{elsarticle-num} 
\bibliography{refs, examples-refs}





\end{document}